\documentclass[12pt]{article}

\usepackage{authblk}
\usepackage[margin=1in]{geometry}
\usepackage[english]{babel} 
\usepackage[utf8]{inputenc} 
\usepackage{graphicx}
\usepackage{natbib}
\usepackage{mathtools}
\usepackage{amsthm}
\usepackage{amsfonts}
\usepackage{amssymb}
\usepackage{bm}
\usepackage{commath} 
\usepackage{mathrsfs}
\usepackage{hyperref}  
\usepackage{cite}
\usepackage[english]{fancyref}
\usepackage{caption}
\usepackage{dsfont} 
\usepackage{tikz}
\usepackage{xcolor}
\usetikzlibrary{bayesnet}
\usepackage[linesnumbered,lined,boxed,commentsnumbered]{algorithm2e}
\usepackage{enumitem}
\usepackage{color, colortbl} 
\usepackage{fancyhdr}
\hypersetup{
	colorlinks=true,
	linkcolor=blue,
	citecolor=blue,
}

\newcommand{\revision}[1]{#1}
\newcommand{\revisiontwo}[1]{#1}

\definecolor{Gray}{gray}{0.9}
\newcolumntype{g}{>{\columncolor{Gray}}c}

\newcommand{\doc}{w}
\newcommand{\metadoc}{\tilde{\X}}
\newcommand{\X}{X}
\newcommand{\W}{X}
\newcommand{\Y}{Y}
\newcommand{\Z}{Z}
\newcommand{\tsum}{\textstyle \sum}
\newcommand{\bigO}{\mathcal{O}}

\DeclareMathOperator{\p}{p} 
\DeclareMathOperator{\q}{\mathcal{R}} 
\DeclareMathOperator{\ICL}{ICL}
\DeclareMathOperator{\Dir}{\mathcal{D}} 
\DeclarePairedDelimiterX{\KLx}[2]{(}{)}{%
	#1\,\delimsize\|\,#2%
}
\DeclareMathOperator{\KLy}{KL}
\newcommand{\KL}{\KLy \KLx}

\DeclareMathOperator{\const}{const}

\newtheorem{prop}{Proposition}

\title{Greedy clustering of count data through a mixture of multinomial PCA}
\author[1,2]{Nicolas Jouvin}
\author[2]{Pierre Latouche}
\author[3]{Charles Bouveyron}
\author[4]{Guillaume Bataillon}
\author[4]{Alain Livartowski}

\affil[1]{Laboratoire SAMM EA 4543, 90 rue de tolbiac, 75013 PARIS}
\affil[2]{Université de Paris, MAP 5, UMR 8145, Paris, France}
\affil[3]{Université Côte d'Azur, Inria, CNRS, Laboratoire J.A. Dieudonné, Maasai team}
\affil[4]{Institut Curie, 25-26 rue d'Ulm 75005 PARIS}

\date{}

\begin{document}
	
	\maketitle

	\begin{abstract}
		Count data is becoming more and more ubiquitous in a wide range of applications, with datasets growing both in size and in dimension. In this context, an increasing amount of work is dedicated to the construction of statistical models directly accounting for the discrete nature of the data. Moreover, it has been shown that integrating dimension reduction to clustering can drastically improve performance and stability. In this paper, \revision{we rely on the mixture of multinomial PCA, a mixture model for the clustering of count data, also known as the probabilistic clustering-projection model in the literature}. Related to the latent Dirichlet allocation model, it offers the flexibility of \textit{topic modeling} while being able to assign each observation to a unique cluster. We introduce a \revision{greedy} clustering algorithm, where inference and clustering are jointly done by mixing a classification variational expectation maximization algorithm, with a branch \& bound like strategy on a variational lower bound. An integrated classification likelihood criterion is derived for model selection, and a thorough study with numerical experiments is proposed to assess both the performance and robustness of the method. Finally, we illustrate the qualitative interest of the latter in a real-world application, for the clustering of anatomopathological medical reports, in partnership with expert practitioners from the Institut Curie hospital. \\
		~ \\
		\textbf{Keywords :} Clustering, Mixture models, Count data, Dimension reduction, Topic modeling, Variational inference
	\end{abstract}

	\section{Introduction}
	
	\subsection{Context}
	Count data is used in many scientific fields in the form of frequency counts for instance in bag-of-words models for text analysis \citep{aggarwal2012survey}, or as next generation sequencing \textit{read} counts in genomics \citep{anders2010differential}. In ecology, a lot of studies also focus on abundance count data \citep{fordyce2011hierarchical}. With the increase in volume and dimensionality of these datasets, there is an interest in summarizing them with the help of new statistical tools, looking for groups of co-expressed genes or meaningful partitions of documents in text corpora. When applied to count data, most of the standard statistical hypothesis acceptable for continuous data, \textit{e.g.} Gaussianity, fall apart. On the one hand, transformations of the raw data have been proposed to meet the normality assumptions, such as log transforms in biology and ecology \citep{zwiener2014transforming, st2018count}, or the well known term frequency-inverse document frequency in text analysis \citep{ramos2003using}. While it is not the purpose of this paper to discuss whether these modifications are statistically well-grounded, we point out the work of \citet{osborne2005notes} and \citet{o2010not}, who emphasized that caution should be taken when using such transformations. On the other hand, statistical model for count data, relying on probabilistic assumptions about the generative process of raw observations, have recently received an increasing amount of attention and developments. \revisiontwo{The goal of this paper is to introduce a new model-based algorithm for count data clustering, capable of handling high-dimensional datasets.}

	\subsection{Model based clustering for count data}
	In an unsupervised setting, clustering consists in looking for a partition of the data in $Q$ groups. Originally treated with distance based methods \citep{hartigan1975clustering}, a flexible statistical framework was then introduced via probabilistic mixture modeling. In model based clustering, a data point is supposed to be drawn from a convex combination of parametric distributions, often called components, with different parameters. Maximum likelihood inference is typically done in the missing data framework of \citet{dempster1977maximum}, where a latent multinomial random variable is assigned to each observation, indicating its component. For a deeper insight on mixture models, we refer to \citet{banfield1993model}, \citet{mclachlan2000finite} and to the recent book of \citet{bouveyron_celeux_murphy_raftery_2019}.
	
	While the Gaussian mixture model constitutes the most popular instance of such probabilistic clustering models, various works extend them to a broader range of distributions, including discrete ones.  In Biology for instance, \citet{rau:hal-01193758} proposed two carefully parameterized Poisson mixture models to cluster RNA-seq count data. The originality of the model being that the Poisson parameters factorize as an individual expression level, a cluster dependent intensity, and an experiment dependent library size. Inference is done by maximizing the complete data log-likelihood, through a classification expectation maximization (CEM) algorithm, introduced in \citet{celeux1992classification}. In a document clustering context, \citet{rigouste2007inference} proposed a detailed evaluation of the multinomial mixture model where components are viewed as multinomial distributions. Comparing an expectation maximization (EM) algorithm with a Gibbs sampler, they obtained comparable performances for both approaches, illustrating the difficulties of high dimensional estimation in document clustering applications with a large vocabulary.  More recently, \citet{silvestre2014identifying} suggested to integrate clustering and model selection in a single algorithm for discrete mixture models. The latter aims at maximizing directly the minimum message length, which is a penalized likelihood criterion, with a modified EM algorithm. 
	
	Again, a drawback of such approaches is that parameter estimation suffers from the dimensionality of the data. This problem is common in many statistical models and roots far beyond discrete models.
	
	\subsection{Dimension reduction}
	Dimension reduction seeks to find an embedding of the data into a lower dimensional subspace. The principal components analysis (PCA) of \citet{hotelling1933analysis} relies on geometrical arguments, searching for linearly uncorrelated pseudo-variables from the original ones. It can also be formulated as a matrix-factorization problem, looking for two low-rank matrices of \textit{loadings} and \textit{scores}, such that their product approximates the data matrix through the euclidean norm \citep{eckart1936approximation}. \citet{tipping1999probabilistic} later drew links with the statistical framework, introducing the probabilistic PCA (pPCA) model where the scores are treated as hidden Gaussian random variables. Inference is done with an EM algorithm. In the last few years, research has focused mainly on finding parsimonious models, in order to tackle high-dimensional problems (see  \textit{e.g.} \citet{mattei2016globally}).
	
	Moving out from the Gaussian setting, several works extended these approaches to a wider range of distributions. \citet{chiquet2018variational} cast pPCA in the generalized linear model framework of \citet{nelder1972generalized}, and then proposed a generalization of pPCA for exponential family link functions, detailing a variational inference procedure for Poisson distributed observations. Non-negative matrix factorization (NMF) algorithms, proposed by \citet{lee2001algorithms}, seek to do matrix factorization with non-negativity constraints and with respect to specific reconstruction errors, such as Euclidean norm or modified Kullback-Leibler divergences. \citet{ding2008equivalence} then showed that the latter formulation may be linked to the probabilistic latent semantic indexing (pLSI) of \citet{hofmann1999probabilistic}, which is a statistical model characterizing the presence of words inside documents. Each word is modeled as a mixture of multinomial components, where the multinomial parameters are discrete distributions over words called \textit{topics}. The document is then represented into this lower-dimensional topic space via its mixture proportions. While inference is conveniently done by an EM algorithm, pLSI lacks a generative process at the document-level, and was shown to be prone to overfitting. 
	
	In order to circumvent this issue, \citet{blei2003latent} proposed a Bayesian formulation of pLSI, called latent Dirichlet allocation (LDA), putting a Dirichlet prior onto the topic proportions for each document, thus making it a fully generative model for new observations. Relying on a fast and efficient variational EM (VEM) algorithm, it soon became a fundamental tool of textual analysis. The dimension reduction aspect of LDA is best understood in his twin formulation called multinomial PCA (MPCA) \citep{buntine2002variational}, drawing a parallel between the topics and latent mixture proportions with the loadings and scores of PCA respectively, thus appearing as a probabilistic matrix factorization method for count data. Together, they form the building blocks of the so-called \textit{topic models}, appearing in a wide variety of domain, such as image analysis \citep{lazebnik2006beyond}, graph clustering \citep{bouveyron2018stochastic} and the analysis of contingency tables \citep{berge2019latent} with textual information.
	
	The key advantage of LDA and MPCA, compared to other models for count data, is their flexibility. In particular, they allow observations to have mixed memberships towards the various topics. As mentioned above, the topic proportions act as lower dimensional representations of the observations \citep{buntine2003multinomial}. In practice, in clustering applications, a simple thresholding of the topic proportions is often not sufficient to retrieve relevant partitions. To tackle this issue, many methods have been considered to post-process the topic proportions using standard clustering algorithms \citep{bui2017combining, liu2016overview}.

	\subsection{Integrating clustering and dimension reduction}
	A considerable amount of works have been dedicated to the construction of models that can take into account the variability in high-dimensional spaces. In the Gaussian setting, \citet{tipping1999mixtures} proposed a mixture of pPCA, later extended in \citet{bouveyron2007high} to account for parsimony. It consists in a Gaussian mixture model where the covariance matrices allow the dimension of the latent subspace to be variable across clusters.
	
	For discrete variables, several works have focused on extending \revision{these ideas to the clustering of count data. Recently, \citet{watanabe2010simultaneous} proposed an extension of the mixture of pPCA to exponential family distributions, putting explicit constraints on their natural parameter. The proposed variational Bayes algorithm relies on iterative clustering-projection phase, where the objective function is a variational lower bound of the model evidence with an additional Laplace approximation step. Specifically relying on topic models,} in Chapter $5$ of her PhD thesis, \citet{wallach2008structured} proposed the cluster topic model (CTM), an extension of LDA, where the latent topic proportions are now drawn from a mixture of $Q$ Dirichlet distributions with different hyper-parameters. Inference is done with a Gibbs sampling algorithm. \revision{\citet{LatentDirichletMixtureModel} proposed a variational Bayes algorithm for inference in the same model, along with a supervised version for text classification.} \citet{xie2013integrating} extended this \revision{model} in their multi-grain clustering  topic model, modeling an observation as a mixture between a \textit{global} and a second mixture of \textit{local} models LDA with different topic matrices. The inference relies on a VEM algorithm. However, we point out that the model is highly parameterized due to the multiple local LDA models parameters, \revision{causing the model to suffer from over-parametrization in high-dimensional problems with few observations.}
	
	\revision{In this paper, we rely on the probabilistic clustering-projection (PCP) model \citep{yu2005probabilistic}, a generative model for count data, relying on MPCA as well as mixture models. In this model, given the latent topic proportions, the law of an observation is a mixture of MPCA with the topics shared across clusters, hence its alternative name: the mixture of multinomial PCA (MMPCA). \citet{yu2005probabilistic} originally proposed a VBEM algorithm for maximum likelihood estimation, then performing clustering with a maximum a posteriori estimates on the posterior cluster membership probabilities.} 
	
	\subsection{Contributions and organization of the paper}
	In this paper, \revisiontwo{we aim at clustering count data in high-dimensional spaces. To this end,} we introduce a greedy inference procedure for MMPCA, focusing on maximizing an integrated classification likelihood. The algorithm is a refined version of the classification VEM (C-VEM) of \citet{bouveyron2018stochastic}, in the spirit of the branch \& bound algorithm, where clustering and inference are done simultaneously. This approach, based on topic modeling, allows to tackle high-dimensional problems, with a limited number of observations. An open-source R package \citep{Rcore} greed that provides a reference implementation of the algorithm introduced in this paper is also available\footnote{\url{https://github.com/nicolasJouvin/MoMPCA}}.
	
	Section~\ref{sec:Section1} presents the model and its characteristics. In Section~\ref{sec:Inference}, the \revision{greedy clustering algorithm} is detailed and a model selection is derived. Then, a thorough study on numerical simulations is detailed in Section \ref{sec:NumericalExperiments}, comparing the performance of MMPCA with other state-of-the-art methods. Finally, Section \ref{sec:RealData} describes a qualitative analysis for the clustering of oncology medical reports, in partnership with two expert doctors, illustrating the capacity of the methodology to uncover useful information from count data.

	\section{The model}
	\label{sec:Section1}
	This section aims at describing the MMPCA model along with notations. In the following, $X = \{x_i\}_{i=1,\ldots, N}$ denotes the set of observations, where $x_i \in \mathbb{N}^V$. The total count for observation $i$ will be noted $L_i := \sum_v x_{iv}$. In text analysis, $V$ denotes the \textit{vocabulary} size when observations are \textit{documents} represented in a \textit{bag-of-words} model, and $x_{iv}$ is the $v$-th word total count inside document $x_i$. In RNA-seq data, $x_{iv}$ represents the total count of reads inside gene $x_i$ in the $v$-th biological sample. In ecology, it might denote the observed number of plants belonging to species $v$ in a geographical site $x_i$. For more details about abundance count data we refer to \citet{cunningham2005modeling}.
	\subsection{Multinomial PCA}
	\label{sec:MPCA}
	A key assumption in probabilistic models for dimension reduction is that each observation $x_i$ can be linked to a latent random variable, that we call $\theta_i$ here, lying in a subspace of dimension $K < V$. The link is generally a combination of a linear transformation $\beta$ on the latent space, and a probabilistic \textit{emission function} parametrized by this transformation. In the probabilistic PCA (pPCA) model introduced in \citet{tipping1999probabilistic}, each $x_i$ lies in $\mathbb{R}^V$ and $\theta_i$ is assumed to be drawn from a standard Gaussian $\mathcal{N}_K(0_K, I_K)$. Then, the conditional law of the observation is again assumed to be Gaussian:
	\[
	x_i \mid \theta_i \sim \mathcal{N}_V(\beta\theta_i + \mu, \sigma^2 I_V) .
	\]
	The model parameters $(\beta, \mu)$ are learned via maximum likelihood inference, as well as the variance $\sigma^2$.
	
\begin{figure}[!ht]
		\centering
		\includegraphics[scale=1]{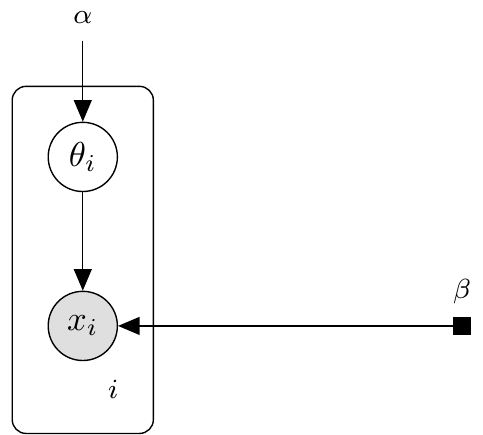}
		\caption{Graphical model of MPCA.}	\label{fig:modelMPCA}
	\end{figure}

	Although the Gaussian hypothesis may make sense for real data, it becomes unrealistic when dealing with non-negative count data. In \citet{buntine2002variational}, the author proposed a discrete analog of pPCA where the latent variables now represent a discrete probability distribution on $\{1,\ldots, K\}$, (\textit{i.e.} $\theta_{ik} \in \Delta_K := \{ p \in \mathbb{R}^K \, : \, p \succcurlyeq 0 \textrm{ and } \sum_{k} p_{k} =1 \}$). Thus, $\theta_i \sim \mathcal{D}_K(\alpha)$ where $\mathcal{D}_K$ is some distribution on $\Delta_K$, almost always chosen to be the Dirichlet distribution: 
	\begin{equation}
	\mathcal{D}_K(\theta_i; \,\mathbf{\alpha}) = \frac{1}{Z(\mathbf{\alpha})} \prod_{k=1}^K \theta_{ik}^{\alpha_k -1} \mathds{1}_{\Delta_K}(\theta_i), \, \textrm{ with} \, \alpha=(\alpha_1, \ldots, \alpha_K)\succcurlyeq 0 .
	\end{equation}	
	Then, the probabilistic emission function is assumed to be multinomial and the model, described in Figure~\ref{fig:modelMPCA}, writes as follow:
	\begin{align}
		\label{eq:MPCA}
		\theta_i  &\sim \mathcal{D}_K(\alpha) , \nonumber \\
		x_i \mid \theta_i &\sim \mathcal{M}_V( L_i, \, \beta \theta_i).
	\end{align}	
	The columns of matrix $\beta \in \mathbb{R}^{V\times K}$ contains $K$ discrete probability distributions on $\{1,\ldots, V\}$, called \textit{topics}. The MPCA model makes the assumption that each observation $x_i$ may be decomposed as a probabilistic mixture of $K$ topics characterizing the whole corpus. Then, an observation is represented by $\theta_i$, the mixture weights in the latent space $\Delta_K$, whereas $\beta$ is a global parameter summing up the information at the corpus level. The \textit{complete} likelihood of $(x_i, \theta_i)$ is then:
	\begin{equation}
	\label{eq:MPCAllhood} 	
	\p(x_i, \theta_i \mid \beta) = \p(\theta_i) \, \frac{L_i!}{\prod_v x_{iv}! \,} \prod_{v=1}^{V}  \left(\beta_{v,\cdot} \theta_i \right)^{x_{iv}} ,
	\end{equation}
	where $\beta_{v,\cdot}$ represents the $v$-th row of $\beta$ as a row-vector. As we will see in Section \ref*{sec:LinkWithLDA}, this model is strongly related to LDA \citep{blei2003latent}, and is the building block for many of the so-called \textit{topic models}. Note that in practice, inference is generally done in the LDA formulation via variational methods \citep[see][for instance]{hoffman2010online}.
	\subsection{Mixture of Multinomial PCA}
	\label{sec:MMPCA}
\begin{figure}[!ht]
		\centering
		\includegraphics[scale=1]{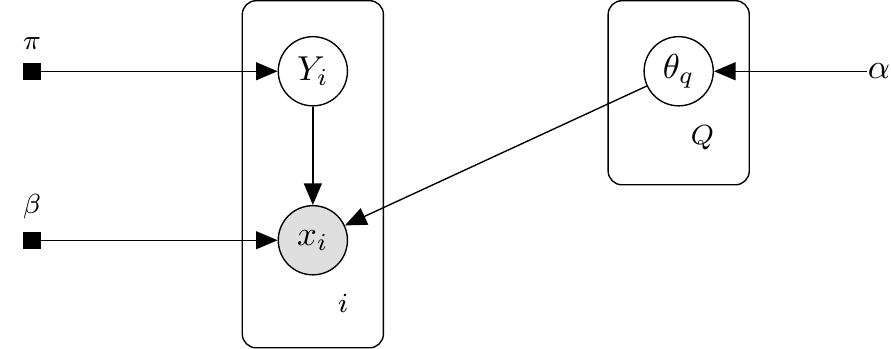}
		\caption{Graphical model of MMPCA.}	\label{fig:modelmomPCA}
	\end{figure}

	Although MPCA allows dimension reduction on discrete data, it is not designed for clustering \textit{per se}. \revision{\citet{yu2005probabilistic} proposed to integrate these two aspects simultaneously, using both topic and mixture modeling, in the same probabilistic model that we call mixture of MPCA (MMPCA) afterwards.} In mixture models with $Q$ components, the cluster assignment of observation $x_i$ is classically represented as a multinomial variable $Y_i \in \{0,1\}^Q$, where $Y_{iq}=1$ if $x_i$ belongs to cluster $q$. We propose a model where $Q$ latent variables are drawn independently:
	\begin{equation}
	\forall q , \, \theta_q \sim \mathcal{D}_K(\,\alpha) .
	\end{equation}  
	Then, conditionally to its group assignment $Y_i$ and the set $\theta=(\theta_q)$, each observation is assumed to follow an MPCA distribution with cluster specific topic proportions:
	\begin{align*}
		\label{model:MMPCA}
		Y_i &\sim \mathcal{M}_K( \, 1,\, \pi), \\
		x_i \mid \{Y_{iq} = 1\}, \, \theta & \sim \mathcal{M}_V( \, L_i, \, \beta \theta_q) . \tag{MMPCA}
	\end{align*}
	The generative model is detailed in Figure~\ref{fig:modelmomPCA}. One of the main difference with MPCA is that the individual latent variable $\theta_i$ now becomes $\theta_q$, at the cluster level, while $\beta$ does not depend on the cluster assignment. Knowing $\theta$, a distribution of interest is the conditional \textit{classification} likelihood, which can be written at the observation level:
	\begin{align}
		\label{eq:completeConditional}
		\p(x_i, Y_i \mid \theta, \, \beta, \pi) &= \p(x_i\mid Y_i, \theta, \, \beta) \, \p(Y_i \mid \pi) , \nonumber \\
		&= \prod_{q=1}^Q \left[\pi_q \, \mathcal{M}_V(x_i; \, L_i, \beta \theta_q) \right]^{Y_{iq}} .
	\end{align}
	Then, marginalizing on $Y_i$ leads to the conditional marginal distribution of an observation:
	\begin{equation}
	\label{eq:marginalConditional}
	\p(x_i \mid \theta, \, \beta, \pi) = \sum_{q=1}^{Q} \pi_q \, \mathcal{M}_V(x_i; \, L_i, \beta \theta_q)  ,
	\end{equation}
	which corresponds to a mixture of MPCA distributions, hence the model name. In the next section, we propose another formulation of the model which will prove useful for inference.
	
	\subsection{Link with Latent Dirichlet Allocation}
	\label{sec:LinkWithLDA}
	As stated above, MPCA is strongly linked to LDA \citep{blei2003latent}. In the context of text analysis, where it was developed, an observation is a document, represented as the set of tokens, or words, $\doc_i = \{\doc_{in},\, n=1,\ldots, L_i\}$ appearing in it, with $\doc_{in} \in \mathbb{N}^V$. Each word $\doc_{in}$ in a document $i$ is first associated with a topic characterized by a vector $z_{in}$ assumed to be drawn from $\mathcal{M}_K(1, \theta_i)$. Then, the word $w_{in}$ is sampled from the distribution $\mathcal{M}_V(1, \beta_{\cdot,k})$, and the model may be written for any document $i$: 
	\begin{align*}
		\label{model:LDA}
		\theta_i & \sim \mathcal{D}_K(\alpha) \\
		\forall n \in \{1,\ldots, L_i\}, \quad z_{in} \mid \theta_i  & \sim \mathcal{M}_K( \, 1, \theta_i) , \nonumber \tag{LDA}  \\
		w_{in} \mid \{ z_{ink}=1 \}  &\sim \mathcal{M}_V(\, 1, \beta_{\cdot,k})  .
	\end{align*}
	At the word-level, marginalizing on $z_{in}$ gives the distribution:
	\begin{equation}
	\label{eq:tokenDistribution}
	w_{in} \mid \theta_i \sim \mathcal{M}_V(1, \beta\theta_i),
	\end{equation}
	which is similar to that of Equation~\eqref{eq:MPCA}. Moreover, it does not depend on the choice of token $n$, thus $(\doc_{in})_n \mid \theta_i$ are independent and identically distributed from Equation \eqref{eq:tokenDistribution}. Hence, the complete likelihood factorizes and can be rearranged as follows:
	\begin{equation}
	\label{eq:LDAllhood}
	\p(w_i, \theta_i \mid \beta) = \p(\theta_i) \, \prod_{n=1}^{L_i}\prod_{v=1}^{V} \left(\beta_{v,\cdot} \theta_i \right)^{w_{inv}} = \p(\theta_i) \, \prod_{v=1}^{V} \left(\beta_{v,\cdot} \theta_i \right)^{\sum_n w_{inv}} .
	\end{equation}
	In MPCA, $x_{iv} = \sum_n w_{inv}$ is the number of time word $v$ of the vocabulary occurred in document $i$. Thus, Equation \eqref{eq:LDAllhood} is almost the likelihood of Equation \eqref{eq:MPCAllhood} except for the missing multinomial coefficient which does not depend on any of the parameters.
	
	Following the reasoning above, a modification of LDA gives an alternative formulation for MMPCA:
	\begin{align*}
		\label{model:mLDA}
		Y_i &\sim \mathcal{M}_Q(1, \, \pi) \nonumber , \\
		\forall n \in \{1,\ldots, L_i\}, \quad z_{in} \mid \{ Y_{iq} = 1\},  \theta_q  &\sim \mathcal{M}_K(1, \, \theta_q) , \tag{MLDA} \\
		w_{in}  \mid \{ z_{ink} = 1 \} &\sim \mathcal{M}_V(1, \, \beta_{\cdot,k})  .\nonumber
	\end{align*}
	Indeed, for any word $w_{in}$, the topic assignment $z_{in}$ can be marginalized out, leaving the distribution: 
	\begin{equation}
	\label{eq:tokenDistributionMLDA}
	w_{in} \mid \{Y_{iq} = 1\}, \theta_q \sim  \mathcal{M}_V(1, \beta \theta_q).
	\end{equation}
	Once again, this distribution is independent of the choice of $n$, hence $(\doc_{in})_n \mid\{Y_{iq} = 1\}, \theta_q$ are independent and identically distributed from \eqref{eq:tokenDistributionMLDA}. Furthermore, the correspondence with MPCA appears clearly when marginalizing on $Y_i$:
	\begin{equation}
	\label{eq:marginalConditional2}
	\p(w_i \mid \theta, \, \beta, \pi) = \sum_{q=1}^{Q}  \pi_q \, \prod_{n=1}^{L_i} \mathcal{M}_V(w_{in}; 1, \beta \theta_q)  =  \sum_{q=1}^{Q} \pi_q \, \prod_{v=1}^{V} (\beta_{v,\cdot} \theta_q)^{x_{iv}}  .
	\end{equation}
	Clearly, Equations \eqref{eq:marginalConditional} and \eqref{eq:marginalConditional2} are equivalent, up to the multinomial coefficients which are independent of the parameters. This equivalence will prove useful in the following, as we will see that it allows to rely on existing inference procedures for LDA. Hence, we will work with the LDA formulation throughout the rest. Note that this implies a slight abuse of notation as $X$ is still employed to design the whole set of observations, regardless of the fact that the token representation $w_i$ is now used.

	\subsection{Construction of the meta-observations}
	While the previous sections discusses some useful properties of MMPCA at an observation level, another interesting feature of the latter arises when working with the whole set of observed variables. Indeed, knowing $\theta$, observations belonging to the same cluster are independent and identically distributed from $\mathcal{M}_V( 1,\, \beta\theta_q)$. This, along with the stability of the multinomial law under addition, suggests an aggregation scheme at the cluster level.
	\begin{prop}[Proof in Appendix \ref{appendix:PreuveMetadoc}] Let $Y = \{Y_1, \ldots, Y_N \}$ be a set of discrete vectors in $\{0,1\}^Q$ characterizing the clustering. Then,
		\label{prop:MetaDocConstruction}
		\begin{equation}
		\label{eq:metadocllhood}
		\p(\W, \theta \mid Y, \, \beta) = \prod_{q=1}^{Q} \left[\p(\theta_q) \, \prod_{v=1}^{V} \left( \beta_{v,\cdot} \theta_q \right)^{ \sum_{i=1}^N Y_{iq} x_{iv}} \right].
		\end{equation}
	\end{prop}
	In the following, we define the aggregated counts of variable $v$ in cluster $q$ as $\metadoc_{qv}(Y) = \sum_{i=1}^N Y_{iq} x_{iv}$. Then, knowing $\Y$, the p.d.f of Equation~\eqref{eq:metadocllhood} is equivalent to that of a LDA model on $Q$ \textit{meta-observations} $\metadoc_q(\Y)$. Therefore, with $Y$ known and fixed, maximum likelihood inference is equivalent in our model with a LDA model on the induced $Q$ meta-observations. Naturally, the construction of meta-observations depends on the clustering $\Y$. In the next Section, we rely on this property and propose a clustering algorithm, alternating between parameter inference in a model with $\Y$ fixed, and a clustering phase where $\Y$ is updated according to the current parameters.

	\section{\revision{A greedy clustering algorithm for MMPCA}}
	\label{sec:Inference}
	We focus in this paper in maximizing the following integrated classification log-likelihood: 
	\begin{equation}
	\label{eq:classificationllhood}
	\log \p(\W,Y \mid \beta,\pi) = \log \sum_\Z \int_{\Theta} \p(\W,\Y, \theta, \Z \mid \beta,\pi) \dif \theta  ,
	\end{equation}
	with respect to the parameters $(\beta, \pi)$ as well as $\Y$. Contrary to the standard missing data framework of \citet{dempster1977maximum}, we emphasize that $\Y$ is not treated as a set of latent variables and the goal is not to approximate its posterior distribution. Conversely, $\Y$ is seen as a set of binary vectors to be estimated through a discrete optimization scheme. Related to \citet{bouveyron2018stochastic}, this approach is grounded on Proposition \ref{prop:MetaDocConstruction} which, conditionally to the knowledge of $\Y$, casts MMPCA as a LDA model with $Q$ meta-observations, for which there exist efficient optimization procedures.
	
	In this section, we propose a classification variational EM (C-VEM) algorithm mixed with an enhanced greedy swapping strategy in order to perform inference and clustering simultaneously . First, we derive a variational bound of Equation \eqref{eq:classificationllhood}, alongside a VEM algorithm for inference. Then, we detail the proposed clustering procedure for the maximization in $\Y$. Finally, a model selection criterion is derived for our model to estimate the number of clusters together with the number of topics, relying on the \textit{integrated} classification likelihood (ICL) of \citet{biernacki2000assessing}.
	
	\subsection{Classification evidence lower bound}
	As discussed above, Equation \eqref{eq:classificationllhood} decomposes as a sum of a LDA term on the $Q$ aggregated meta-observations, plus a clustering term as follows:
	\begin{equation}
	\label{eq:sumOfTwoTerms}
	\log \p(\W,Y \mid \beta,\pi) = \log \sum_\Z \int_{\Theta} \p\left(\metadoc(Y), \theta, \Z \mid \Y, \beta\right) \dif \theta \, + \,  \log \p\left(\Y \mid \pi\right). 
	\end{equation}
	Here, $\metadoc(\Y)$ represents the collection of the $Q$ meta-observations $(\metadoc_q(\Y))_q$. Unfortunately, neither the integral in Equation \eqref{eq:sumOfTwoTerms}, nor the posterior distribution of latent variables $\p(\Z, \theta \mid \Y, \W, \, \beta, \pi)$ have any analytical form. To tackle this issue, we propose to resort to variational approximation. Introducing a distribution $\q(Z, \theta)$ on the latent variables, the following identity is true, for any clustering $\Y$:
	\[
	\log\p(\X, \Y \mid \pi, \beta) = \mathcal{L}(\q(\cdot); \, \pi, \beta, \Y)  \, + \, \KL{\q(\cdot)}{\p(\cdot \mid \W, \Y, \pi, \beta)} ,
	\]
	with
	\begin{equation}
	\label{eq:CELBO}
	\mathcal{L}(\q(\cdot); \, \pi, \beta, \Y) = \mathbb{E}_{\q}\left[ \log \frac{\p(\W, \Y, \Z, \theta \mid \pi, \beta)}{\q(\Z, \theta)}\right] .
	\end{equation}
	Here $\KLy$ denotes the Kullback-Leibler divergence between the variational distribution $\q(\cdot)$ and the posterior $\p(\cdot \mid \W,\Y, \,\pi, \beta)$:
	\[
	\KL{\q(\cdot)}{\p(\cdot \mid \W, \Y, \pi, \beta)} = - \sum_{\Z} \int_{\theta} \q(Z,\theta) \log\, \frac{\p(Z, \theta \mid \W, \Y,\, \pi, \beta)}{\q(Z,\theta)}.
	\]
	Since the latter is always positive, Equation \eqref{eq:CELBO} constitutes a lower bound of the integrated classification log likelihood, which is an analog of the evidence lower bound in the standard VEM framework. Furthermore, following \citet{blei2003latent}, we assume that $\q(\cdot)$ factorizes over the two sets of latent variables, \textit{i.e.}:
	\[
	\q(\Z, \theta) = \prod_i \prod_n \q(z_{in}) \prod_q \q(\theta_q) .
	\]
	
	\subsection{Optimization}
	\label{subsec:Optimization}
	Considering $Y$ fixed for now, the goal is to maximize $\mathcal{L}$, with respect to $\q(\cdot)$ and the parameters $(\pi,\beta)$. We consider a coordinate ascent, cycling over $\q$ and $(\pi, \beta)$, while maintaining one fixed. Indeed, the objective can easily be rewritten as the sum of a LDA bound on the $Q$ meta-observations and a clustering term.
	\begin{prop}[Proof in Appendix \ref{appendix:PreuveDecomposeBound}]
		\label{prop:DecomposeBound}
		\begin{equation*}
			\label{eq:decomposition}
			\mathcal{L}(\q(\cdot); \, \pi, \beta, \Y) = \mathcal{J}_{\textrm{LDA}}(\q(\cdot); \, \beta, \Y) \, + \, \log \p(\Y \mid \pi) ,
		\end{equation*}
		where
		\begin{equation}
		\label{eq:ELBO}
		\mathcal{J}_{\textrm{LDA}}(\q(\cdot); \, \beta, \Y) = \mathbb{E}_{\q}\left[ \log \p (\metadoc(Y), \Z, \theta \mid \Y, \beta)\right] - \mathbb{E}_{\q}\left[\log \q(Z, \theta) \right] .
		\end{equation}
	\end{prop}
	With such a decomposition, maximizing $\mathcal{L}$ with respect to $\pi$ is direct, and most of the work lies in the maximization of $\mathcal{J}_{\textrm{LDA}}$ with respect to $\beta$ as well as $\q$. The latter can efficiently be done by constructing the meta-observations $\metadoc(Y)$ and using the VEM algorithm of \citet{blei2003latent}.
	
	The following propositions detail the update for each individual distribution, \textit{i.e.} the so-called VE-step obtained from the maximization of Equation \eqref{eq:ELBO}.
	\begin{prop}[Proof in Appendix \ref{appendix:PreuvePhi}]
		\label{prop:PhiUpdate}
		The VE-step update for $\q(z_{in})$ is given by:
		\[
		\q(z_{in}) = \mathcal{M}_K(z_{in};\,  1, \,\phi_{in} = (\phi_{in1}, \ldots, \phi_{inK})),
		\]
		with
		\[
		\forall (i, n, k), \quad \phi_{ink} \propto \left( \prod_{v=1}^V \beta_{vk}^{w_{inv}} \right) \, \prod_{q=1}^Q \exp\left\{ \psi(\gamma_{qk}) - \psi\left(\textstyle \sum_{l=1}^K \gamma_{ql}\right)\right\}^{Y_{iq}} .
		\]
	\end{prop}
	\begin{prop}[Proof in Appendix \ref{appendix:PreuveGamma}] 
		\label{prop:GammaUpdate}
		The VE-step for $\q(\theta)$ is 
		\[
		\q(\theta) = \prod_{q=1}^{Q} \Dir_K\left(\theta_q; \, \gamma_q=(\gamma_{q1}, \ldots, \gamma_{qK})\right) ,
		\]
		with
		\[
		\forall (q,k), \quad \gamma_{qk} = \alpha_k + \sum_{i=1}^{N} Y_{iq}\sum_{n=1}^{L_i} \phi_{ink} .
		\]
	\end{prop}
	\noindent A fixed point algorithm is used, alternating between updates of Propositions \ref{prop:PhiUpdate} and \ref{prop:GammaUpdate}, until the bound converges. Regarding $(\pi, \beta)$, they appear in separate terms of $\mathcal{L}$. The maximization with respect to $\beta$ corresponds to the M-step maximizing Equation \eqref{eq:ELBO}, whereas the optimal $\pi$ is simply the standard mixture proportion estimate.
	\begin{prop}[Proof in Appendix \ref{appendix:PreuveBeta} and \ref{appendix:PreuvePi}]
		\label{prop:BetaAndPi}
		The M-step estimates of $\beta$ and $\pi$ respectively are:
		\begin{align*}
			\forall (v,k) , \quad & \beta_{vk} \propto \sum_{i=1}^{N} \sum_{n=1}^{L_i} \phi_{ink} \, w_{inv} ,\\
			\forall q, \quad &\pi_q \propto \textstyle \sum_{q=1}^Q Y_{iq}.
		\end{align*}
	\end{prop}
	\noindent We now detail a clustering algorithm for MMPCA to estimate $\Y$.
	
	\subsection{A clustering algorithm for MMPCA}
	Optimizing the lower bound $\mathcal{L}$ in $\Y$ is a combinatorial problem, involving to search over $Q^N$ possible partitions. Although it is not possible to find a global maximum within a reasonable time, several heuristics have been proposed to explore efficiently local maxima. Among them, greedy methods have received an extended amount of attention. \revision{Notably, \citet{bouveyron2018stochastic} proposed a C-VEM algorithm for the clustering of nodes in networks. While applicable in this setting, a regular C-VEM algorithm converges to local maxima of the variational lower bound leading to poor clustering performances. Hence,} we propose a refined version of the C-VEM algorithm inspired from the branch \& bound methods. Considering an initial clustering solution $\Y$, the algorithm starts by the VEM of Section \ref{subsec:Optimization}, with $\Y$ fixed, and then cycles randomly through the observations. For each $x_i$, all possible cluster swaps are tested, modifying $Y_i$, and leaving other observations unchanged. For each swap, meta-observations are updated and the VEM algorithm above is used again to update the variational distributions and the parameters. Then, the swap inducing the greatest positive variation of $\mathcal{L}$ is validated, if any, and $(\Y, \pi, \beta, \q)$ are updated accordingly. Moving to the next observation, the algorithm repeats the procedure until no possible swaps increasing the bound may be found, or when a user-defined maximum number of iterations is reached. The whole procedure is described in Algorithm \ref{alg:BranchAndBound} as a pseudo-code. A key difference between the C-VEM algorithm of \citet{bouveyron2018stochastic} is that parameters and variational distributions are updated for each swaps in the greedy procedure, instead of being held fixed. This strategy is close to a \textit{branch} \& \textit{bound} procedure, the lower bound acting as the surrogate for the objective
	\[
	\forall (\Y, \q, \pi, \beta), \quad \log \p(\W, \Y \mid \pi, \beta) \geq \mathcal{J}_{\textrm{LDA}}(\q(\cdot), \, \beta, \Y) \, + \, \log \p(\Y \mid \pi) ,
	\]
	the goal is to efficiently explore a part of the decision tree by temporarily validating a swap, constructing new meta-observations, and re-maximizing the bound with respect to the parameters. It can be done efficiently thanks to the fact that a given swap, from cluster $l$ to cluster $q$, only affects meta-observations $\metadoc_l$ and $\metadoc_q$. Thus, the cost of each VE-step is considerably reduced since the only needed updates concern observations in these two clusters.
	\SetKw{KwBreak}{Break}
	\newsavebox{\VEM}
	
	\begin{algorithm}[ht!]
		\begin{lrbox}{\VEM}
			\verb!VEM!
		\end{lrbox}
		\caption{Branch and Bound C-VEM algorithm}
		\label{alg:BranchAndBound}
		\KwData{$\X$}
		\KwResult{Clustering $\Y$}
		\KwIn{$Q$, $K$, any initializations for $\Y$ and $\beta$. Maximum number of epochs: $T$.}
		\BlankLine
		$\mathcal{L} \gets$ \usebox{\VEM}$(\X, \Y)$
		\BlankLine
		\For{$t \gets 1$ \KwTo $T$}{
			$Y^{(old)} \gets Y$
			\BlankLine
			\For{$i \gets 1$ \KwTo $N$}{
				Find $l$ such that $ Y_{il} = 1$  \\
				\BlankLine
				\For{$q \gets 1$ \KwTo $Q$}{
					\eIf{ $q \neq l$ }{
						Set $Y_{iq}^{(tmp)}=1$ and \\
						$\mathcal{L}[q] \gets$ \usebox{\VEM}$(\X, \Y^{(tmp)})$ \\
						
						Compute:
						$\Delta_i(l, q) \gets \mathcal{L}[q] - \mathcal{L}$ .

					}{
					$\Delta_i(l, q) \gets 0 $
				}
			}
			\BlankLine
			$ q^{\star} \gets \arg\max_q  \Delta_i(l, q) $  \\
			\BlankLine
			\uIf{$ q^{\star} \neq l$ and $C_{l} > 1$}{
				Set $Y_{iq^\star}  = 1$, and $\mathcal{L} \gets \mathcal{L}[q^\star]$ \\
			}
		}
		\lIf{$Y == Y^{(old)}$}{ \KwBreak }
	}
\end{algorithm}
Both VEM and greedy procedures are only ensured to converge to local maxima of $\mathcal{L}$, and we recommend several restarts with different initial clustering solution $\Y$, selecting the run achieving the greatest value. We also found that $\beta$ plays a crucial role in the optimization algorithm. Therefore, we recommend to estimate it with a regular LDA on the whole set of observation at the beginning, without aggregating it, and to use it as a starting value for $\beta$. \revision{Regarding the initialization of $\Y$, we found that there is a negligible impact of using a refined initialization strategy instead of a random balanced one. The methodology is robust to the initialization strategy, which is due to the ability of the branch \& bound approach to efficiently explore the space of partitions.}

\subsection{Model selection}
\label{sec:ModelSelection}
So far, everything described above considered the number of clusters $Q$ and topics $K$ given and fixed. Thus, we still need to handle the task of estimating the best pair $(Q,K)$, which can be viewed as a model selection problem. Several criteria have been proposed for this task, most of them relying on a penalized marginal log-likelihood such as the Akaike information criterion \citep[AIC]{akaike1998information}, or the Bayesian information criterion \citep[BIC]{schwarz1978estimating}. In \citet{carel2017simultaneous}, such criteria are proposed for a frequentist version of MMPCA, where the marginal likelihood is maximized directly. In a clustering context, working with a classification likelihood, \citet{biernacki2000assessing} proposed the ICL criterion for Gaussian mixtures. Following this work, we propose a ICL-like criterion for our model, designed to approximate the likelihood of Equation \eqref{eq:classificationllhood} integrated with respect to the parameters: $\log \p(\W, \Y)$. The proposition hereafter results from a Laplace approximation combined with a variational estimation of the maximum log-likelihood, alongside a Stirling formula on the marginal law of $\Y$.
\begin{prop}[Proof in Appendix \ref{appendix:PreuveICL}] A ICL criterion for MMPCA can be derived
	\label{prop:ICL}
	\begin{eqnarray}
	\label{eq:ICL}
	\ICL_{MMPCA}(Q, K ) & = & \mathcal{L}^\star\left(\q(\cdot); \, \pi, \beta, \Y\right) \nonumber \\
	& & - \frac{K (V-1)}{2} \log(Q) - \frac{Q-1}{2} \log(N) ,
	\end{eqnarray}
	where $\mathcal{L}^\star$ is the lower bound evaluated after convergence of Algorithm \ref{alg:BranchAndBound}.
\end{prop}

\subsection{Run time and complexity}
\label{subsec:Complexity}
We now detail the algorithmic complexity of one epoch of Algorithm \ref{alg:BranchAndBound}, where $\beta$ is initialized once at the beginning and fixed. For an arbitrary observation $x_i$ belonging to cluster $l$, all possible $Q-1$ swaps from cluster $l$ to cluster $q$ are tested, where each swap has the computational cost of two VE steps in LDA. Indeed, from an implementation point-of-view, the only meta-observations affected by the swap are $\metadoc_l(\Y)$ and $\metadoc_q(\Y)$. Hence, we just need to update these two meta-observations accordingly, and run the VE-step described in \citet{blei2003latent} on it. The latter is simply the cost of computing $(\phi_{l} , \phi_{q})$ and $(\gamma_{l} , \gamma_{q})$ which is $\bigO(VK)$. Indeed $(\phi_{l} , \phi_{q} )$ requires to compute $2KV$ coefficients, whereas $(\gamma_{l} , \gamma_{q})$ requires only $2K$. There is an alternation between these two steps until convergence of the evidence lower bound, but, in practice, the convergence is really fast and there is no need for more than a few iterations for each VE-step. In conclusion, it makes $\bigO(NQKV)$ operations for one epoch.
In the experimental setting of Section \ref{subsec:RobustnessToNoise}, one run of Algorithm \ref{alg:BranchAndBound} takes between $2$ and $3$ min on a single CPU with a frequency of $2.3$ GHz\revisiontwo{, and Figure~\ref{fig:ComputationalTime} shows the computational time evolution according to $N$.}

\revisiontwo{Regarding the amount of memory required to store the distribution $\mathcal{R}$ and the parameters $(\pi, \beta)$, Algorithm~\ref{alg:BranchAndBound} (or a regular C-VEM) requires $O(QK + KV + QVK)$ of memory space to store those elements. It is worth noticing that this quantity is constant regarding the number of observations $N$.}

\subsection{Related work} 

Recently, \citet{carel2017simultaneous} proposed the NMFEM algorithm for maximum likelihood inference in a frequentist version of our model. Both generative models are essentially the same except that the cluster latent variables $\theta = (\theta_q)_q$ are now viewed as parameters. However, the inference and optimization procedures differ, since the authors propose to focus on a marginal likelihood maximization through a regular EM algorithm. In this formulation, the E-step consists in computing the posterior distribution $\p(\Y \mid \X, \theta, \beta, \pi)$ which is available in closed form, not relying on variational approximations. As for the M-step, the authors proposed to rely on the multiplicative updates of \citet{lee1999learning} in order to maximize the EM lower bound with respect to $\theta$ and $\beta$ iteratively. Clustering is done using a MAP estimate on the posterior of $\Y$ after convergence. The numerical performances of both models are compared on simulated datasets in the following Section, along with other count data clustering methods.

\section{Numerical Experiments}
\label{sec:NumericalExperiments}
A specific simulation scheme is detailed in the following, in order to evaluate the performance of Algorithm~\ref{alg:BranchAndBound}. 
\subsection{Experimental setting}
Hereafter, unless stated explicitly otherwise, the number of observation is fixed to $N=400$, with total count $L_i=250, \, \forall i$. The matrix $\beta$ is computed once and only on the whole corpus with a mixed strategy of a Gibbs sampling estimate as a starting point for the VEM algorithm of \citet{blei2003latent}. The maximum number of epoch in Algorithm \ref{alg:BranchAndBound} is fixed to $T=7$, and $Y$ initialized randomly. 

We describe hereafter how we simulate data from an MMPCA model. 
We propose to use the following values for model parameters:
\[ 
Q^\star = 6, \, K^\star=4, \,\theta^\star = \begin{bmatrix}
0.50 & 0.17 & 0.17 & 0.17 \\ 
0.17 & 0.50 & 0.17 & 0.17 \\ 
0.17 & 0.17 & 0.50 & 0.17 \\ 
0.17 & 0.17 & 0.17 & 0.50 \\ 
0.33 & 0.17 & 0.33 & 0.17 \\ 
0.17 & 0.33 & 0.17 & 0.33 \\ 
\end{bmatrix} .
\]
It corresponds to a setting where each of the four first clusters are peaked towards one of the four topics, whereas the last two clusters are more \textit{mixed} across topics. 

\begin{figure}[ht!]
	\centering
	\includegraphics[width=\textwidth]{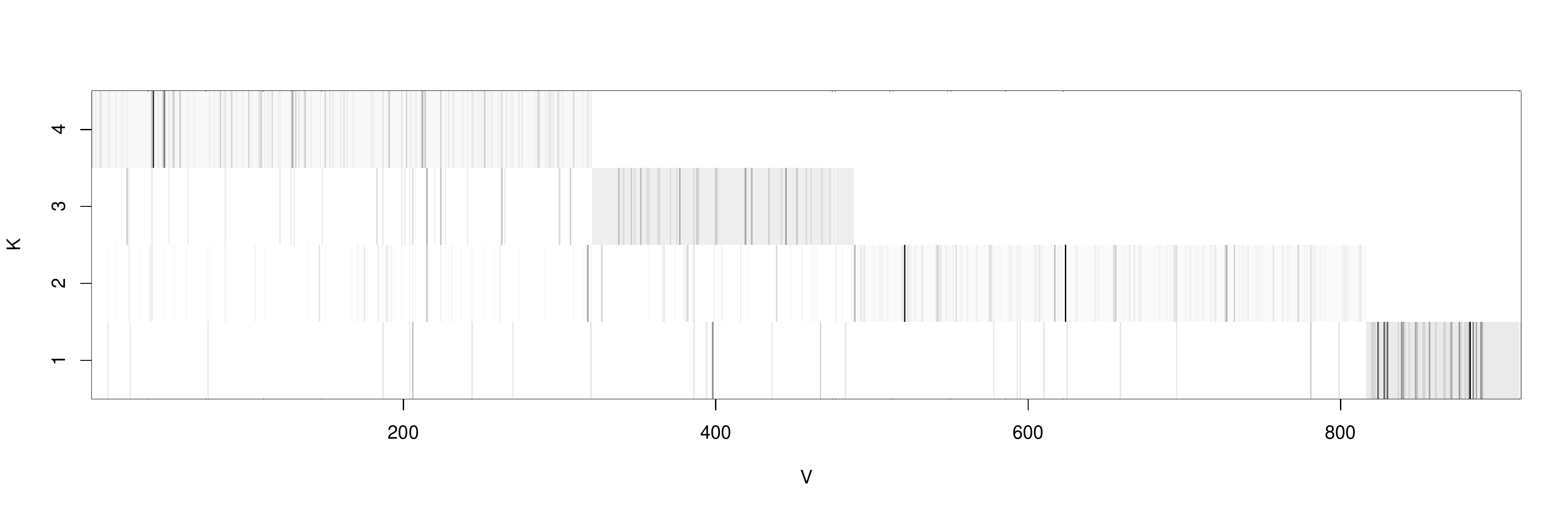}
	\caption{Visualization of the matrix $\beta^\star$. Darker grey indicates stronger probabilities.}
	\label{fig:BetaStar}
\end{figure}

Topics are defined using the empirical distribution of words across four different articles from BBC news, talking about unrelated issues: the birth of princess Charlotte, black holes in astrophysics, UK politics, and cancer diseases in medicine. The matrix $\beta^\star$ is then simply computed as the row-normalized document-term matrix of those four messages, and exhibits a strong block structure, implying that each topic uses a different set of words, as shown in Fig. \ref{fig:BetaStar}. The vocabulary size is $V=915$, which makes it a fairly high dimensional problem. 

As we are dealing with a clustering task, a similarity metric, invariant to label switching, should be used to evaluate the quality of the recovered partition. Several choices are possible in the literature, here we chose the \textit{Adjusted Rand Index} of \citet{rand1971objective} as it is a widely used and accepted metric in the clustering literature.

All experiments were run using the \verb|R| programming language with  the following methods comparison:
\begin{enumerate}[itemsep=0pt]
	\item The non-negative matrix factorization algorithm proposed in \citet{xu2003document}, denoted as \verb|NMF|.
	\item \revision{A clustering found by maximum a posteriori on the latent topic proportions of a LDA model. Inference is done with a VEM algorithm, with $K$ fixed to $Q^\star$.}
	\item  A Gaussian mixture model (GMM) with $Q^\star$ components in the latent space $\theta$ of an LDA with $K^\star$ topics. This method will be called \verb|GMM.LDA|.
	\item A simple mixture of multinomial model for count data clustering, denoted as \verb|MixMult|.
	\item The \verb|NMFEM| algorithm of \citet{carel2017simultaneous} which is another inference procedure for a frequentist version of MMPCA where $\theta$ is treated as a parameter. 
	\item A specific Poisson mixture model for the clustering of high-throughput sequencing data proposed in \citet{rau:hal-01193758}. This method is denoted as \verb|HTSclust|, from the eponym package.
\end{enumerate}
Our implementation of Algorithm~\ref{alg:BranchAndBound} also relies on the \verb|topicmodels| package of \citet{hornik2011topicmodels}\footnote{Available on the CRAN} for the VE-steps and lower bound computation detailed in Section \ref{subsec:Optimization}

\subsection{An introductory example}

\begin{figure}[!ht]
	\centering
	\includegraphics[width=1\textwidth]{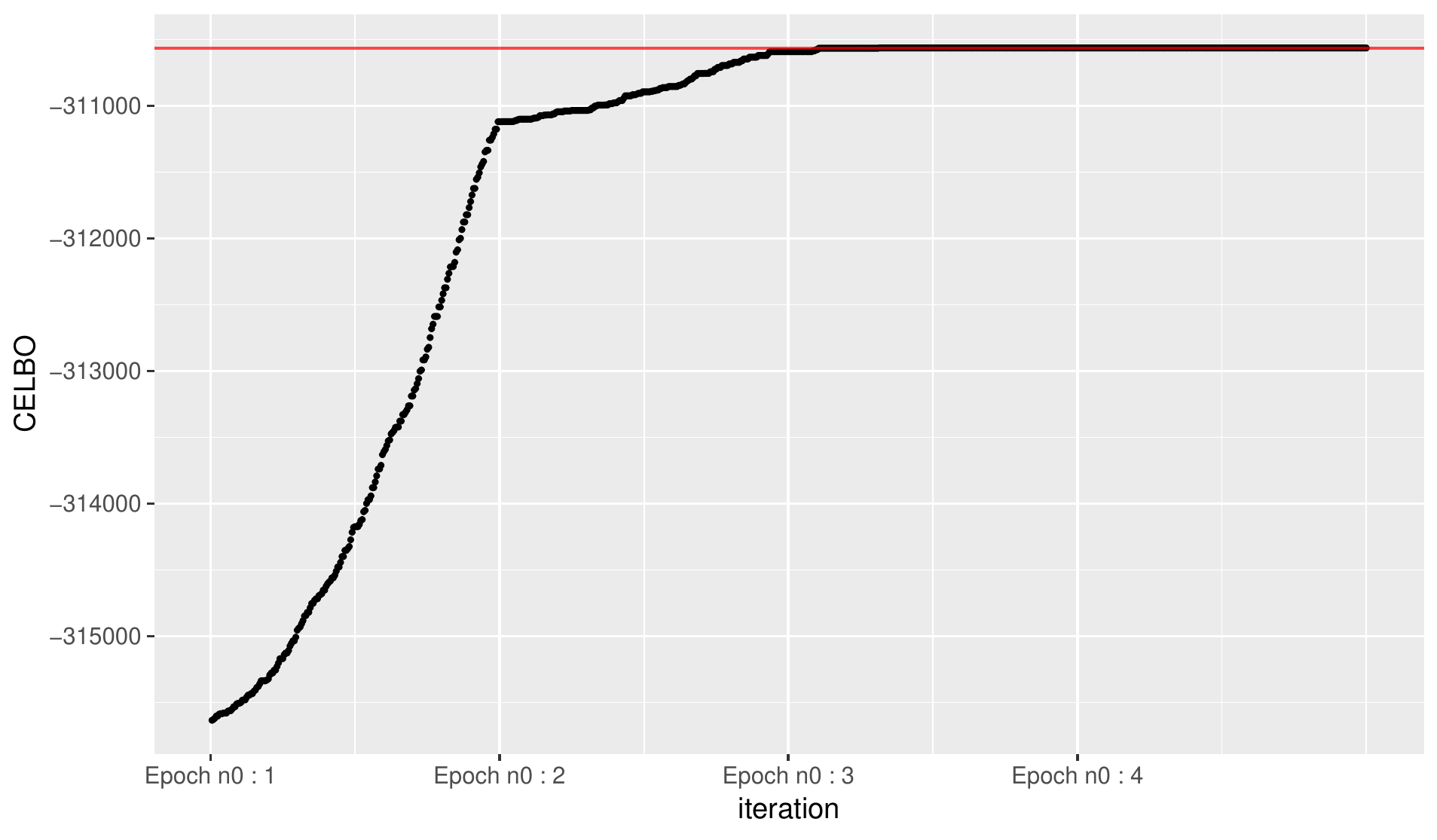} \hfill
	\includegraphics[width=1\textwidth]{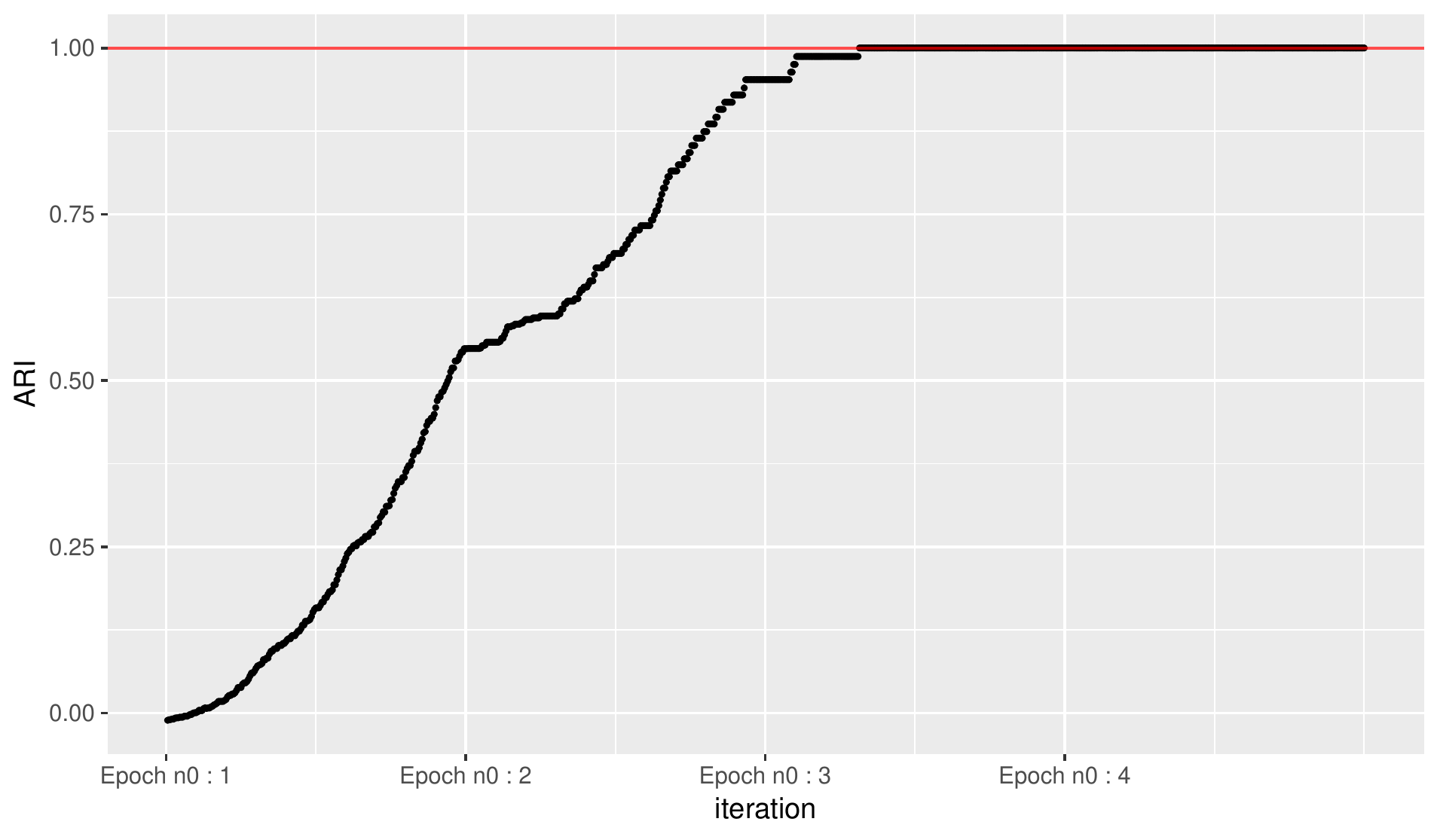}
	\caption{Lower bound (up) and ARI (down) evolution during a full run of the algorithm.}
	\label{fig:ari_and_celbo_evolution}
\end{figure}

Figure~\ref{fig:ari_and_celbo_evolution} shows the joint evolution of the variational bounds and the adjusted rand index on a run of Algorithm \ref{alg:BranchAndBound}. The random initialization gives an ARI close to $0$, which is expected, then we observe a quick maximization of the bound on first epoch, which also corresponds to an amelioration of the ARI. After the first epoch, the bound growth is less pronounced, although swaps still happen at this stage. It tends to indicate that the marginal bound increase of a swap is decreasing. Furthermore, the passage from a good partition to the true one is done with an almost constant bound in the third epoch. Once the true partition is attained, no more swaps can maximize the bound. Hence, in this simple setting, the local maxima of the bound coincide with a maximum ARI. In the next section we propose more complex simulations through the addition of a noise parameter.

\subsection{Robustness to noise}
\label{subsec:RobustnessToNoise}
Leaving $\beta^\star$ unchanged, hence controlling for its complexity, we propose to focus on $\theta^\star$ to investigate the robustness of our method. Indeed, in order to complicate the simulation, we introduce noise in the observations by changing the distribution in the latent space. Indeed, fixing $\epsilon \in [0,\,1]$ and modifying the generative process of the MMPCA model described in \ref{sec:LinkWithLDA}, we now draw: 
\[
z_{in} \mid \{Y_{iq}=1\}, \theta^\star_q  \, \sim \, (1 - \epsilon) \, \mathcal{M}_K(1, \theta^{\star}_{q}) \, + \, \epsilon \, \mathcal{U}(\{1,\ldots,K\})
\]
Thus, $\epsilon = 0$ implies that each token in cluster $q$ follows the standard MMPCA distribution $\mathcal{M}_K(1, \theta^{\star}_{q})$. \revision{When $\epsilon$ reaches $1$, there is absolutely no cluster structure to be found and the groups are totally mixed since they all share the same common discrete distribution over topics $\mathcal{U}(\{1,\ldots, K\})$.}

Moreover, the strength of mixture modeling approaches is also to capture unbalanced cluster sizes. We propose to control group proportions via a parameter $\lambda$ such that $\pi_q \propto \lambda^{Q^\star - q}$. The case $\lambda=1$ corresponds to balanced clusters, whereas $\lambda < 1$ put more emphasis on cluster $5$ and $6$, which may be considered as the \textit{difficult} ones, considering that they are peaked towards two topics instead of only one.

\begin{figure}[!ht]
	\centering
	\includegraphics[width=\textwidth]{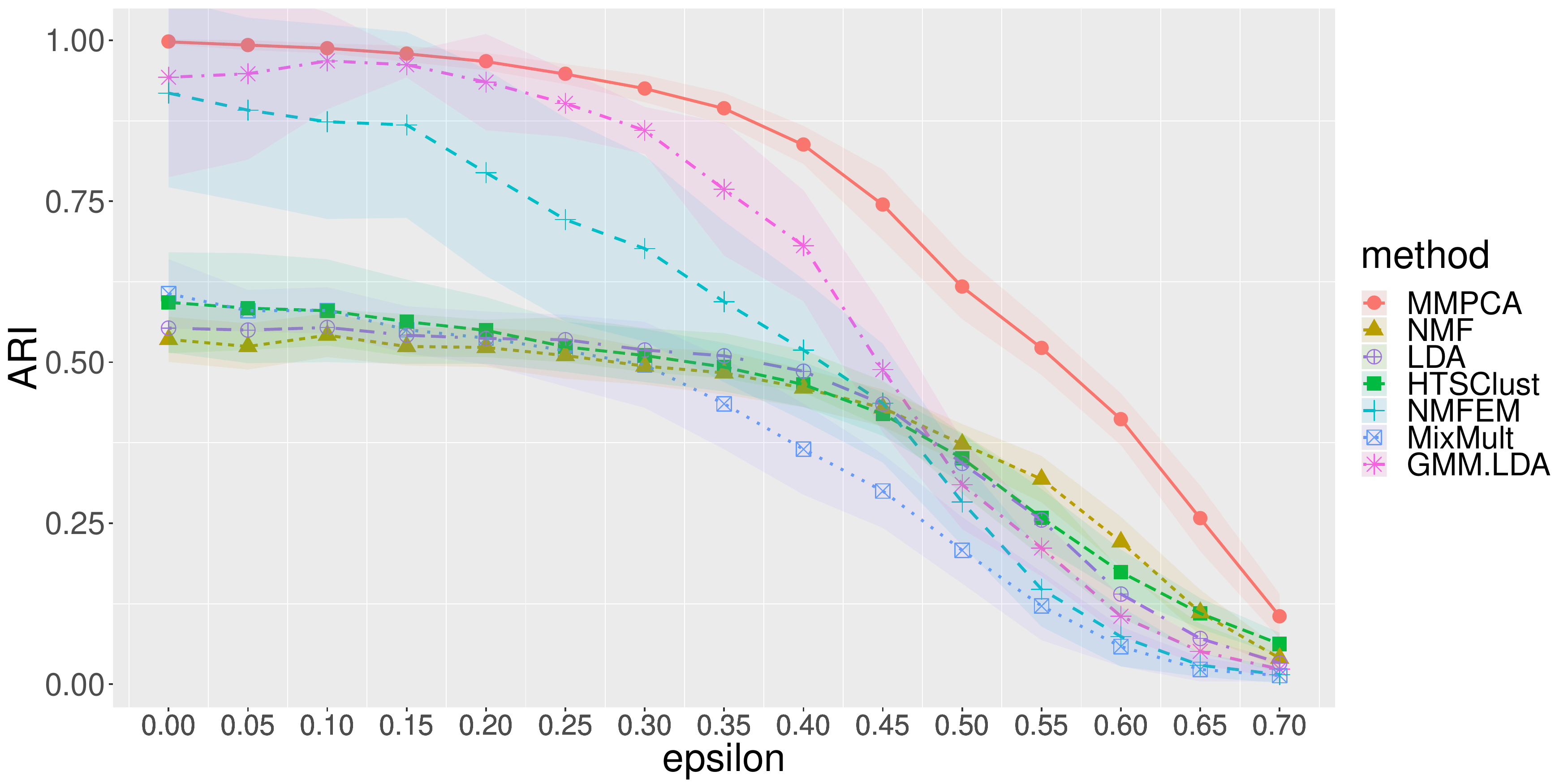} \hfill
	\caption{$\lambda=1$. Mean ARI per noise level $\epsilon$, with error bars. Each score is calculated on \revision{$50$ simulated datasets}.}
	\label{fig:ari_noise_lambda1}
\end{figure}

\begin{figure}[!ht]
	\centering
	\includegraphics[width=\textwidth]{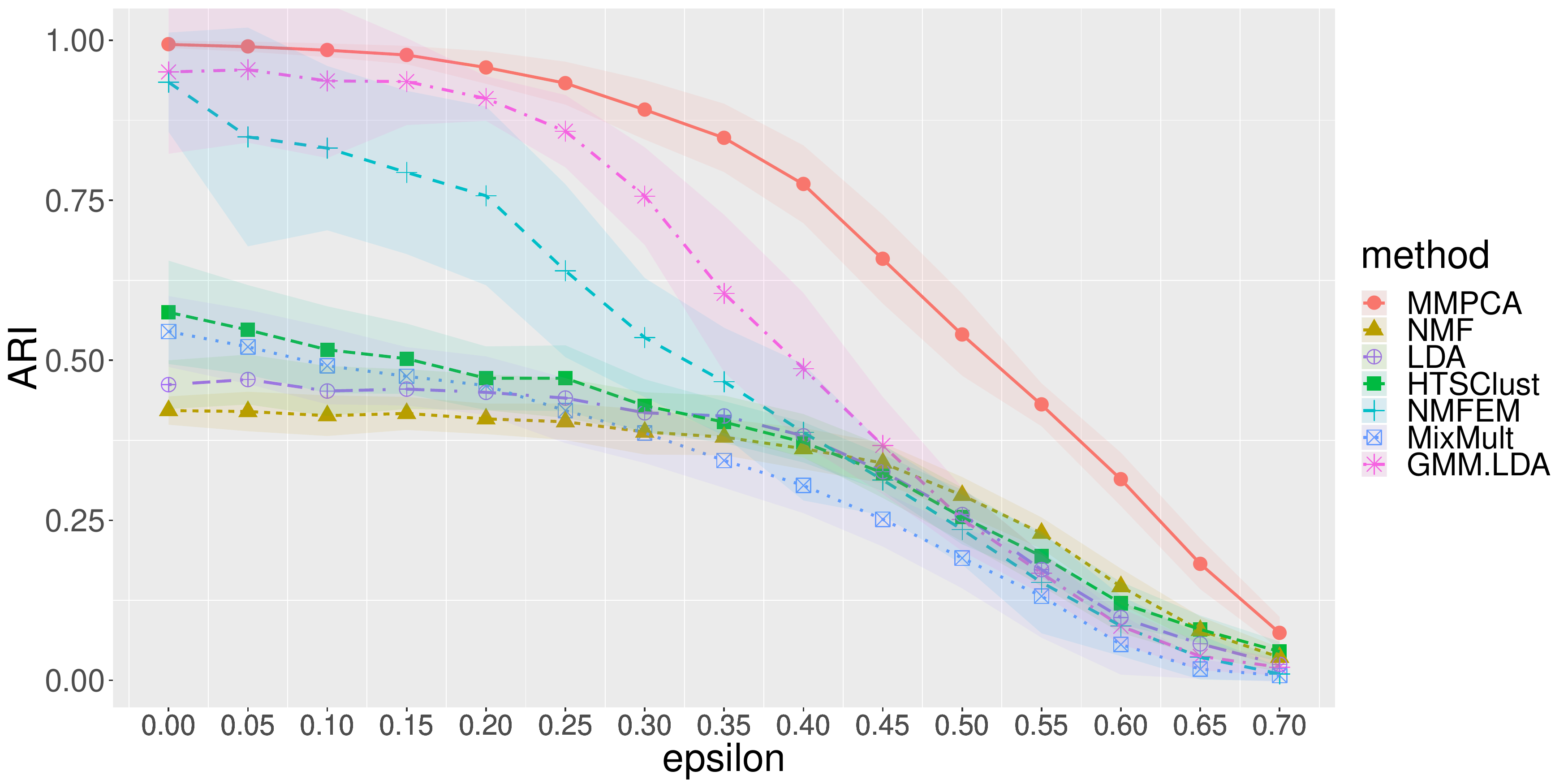} \hfill
	\caption{$\lambda=0.85$. Mean ARI per noise level $\epsilon$, with error bars. Each score is calculated on \revision{$50$ simulated datasets}.}
	\label{fig:ari_noise_lambda085}
\end{figure}

\begin{figure}[!ht]
	\centering
	\includegraphics[width=\textwidth]{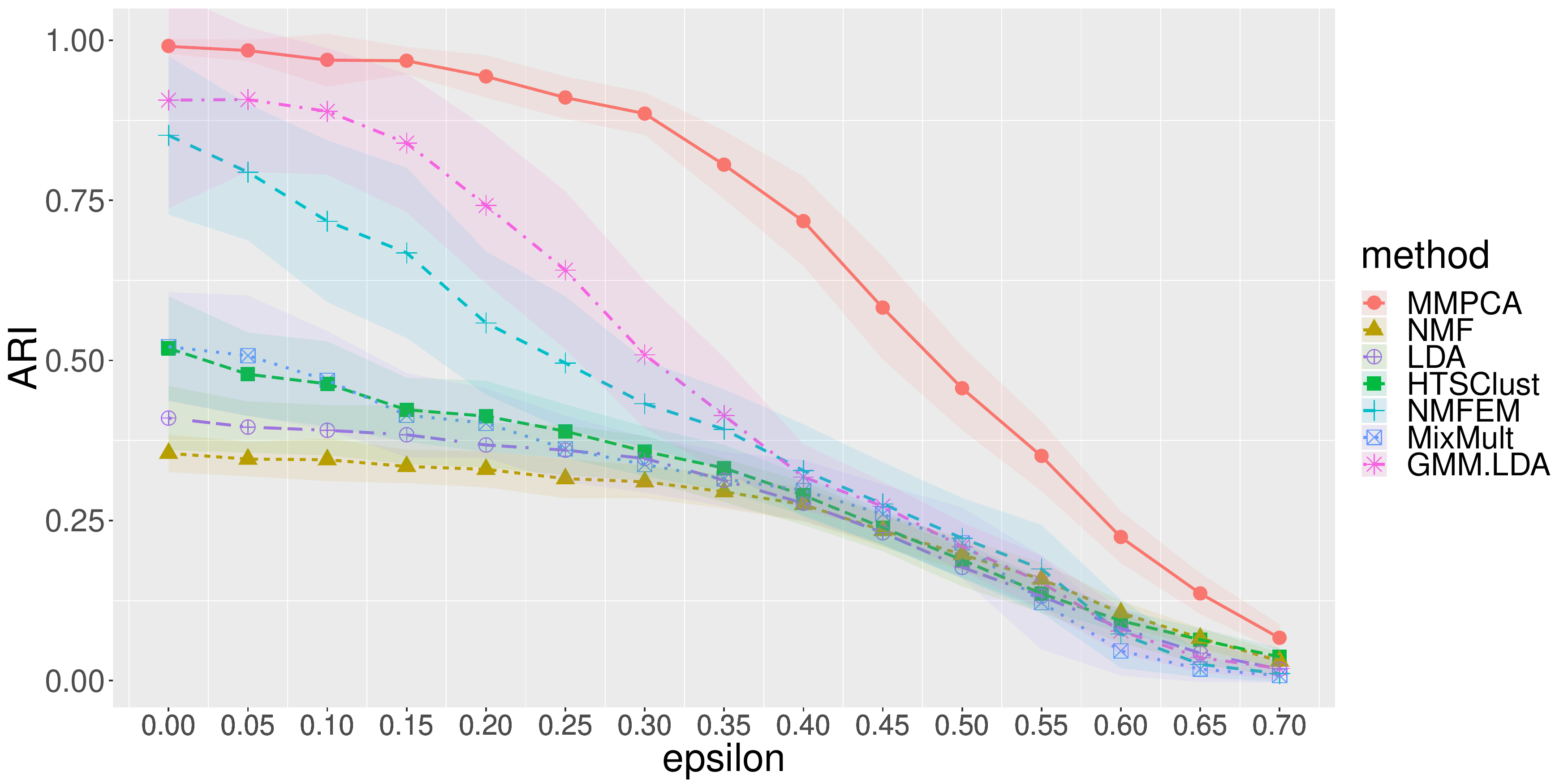} \hfill
	\caption{$\lambda=0.7$. Mean ARI per noise level $\epsilon$, with error bars. Each score is calculated on \revision{$50$ simulated datasets}.}
	\label{fig:ari_noise_lambda07}
\end{figure}

Figure \ref{fig:ari_noise_lambda1}, \ref{fig:ari_noise_lambda085} and \ref{fig:ari_noise_lambda07} represent the mean ARI of each method with respect to the noise level, for $\lambda=1$, $0.85$ and $0.7$ respectively. For every possible pair $(\lambda, \epsilon)$, means and standard errors are computed across \revision{$50$ simulated datasets}. The noise grid goes from $\epsilon=0$, by $0.05$ steps, to $\epsilon=0.7$, since beyond this limit none of the tested methods is able to recover the true partition, the cluster structure behind being almost non-existent.

Overall, MMPCA performs really well when compared to competitors, demonstrating a robustness both to noise and unbalanced clusters. The best competitor seems to be \verb|GMM.LDA|, which, while basic, is advantaged by the knowledge of $(Q^\star, K^\star)$, despite lacking of a model selection criterion. The NMFEM method, which is the closest to our model, seems to perform quite correctly for low noise levels, but exhibits poor stability and efficiency with respect to noise. Moreover, it really seems to suffer from the high-dimensional setting, with fewer observations than variables. The stability of MMPCA over NMFEM advocates for the Bayesian approach, putting a prior on $\theta$, which allow to smooth the dimensionality effect. The differences may also arise from the marginal versus classification likelihood maximization, and the algorithms used for optimization. The mixture of multinomial is really sensitive to noise and to the high-dimensional setting as well, thus supporting the idea of a latent topic factorization of the true parameters. \revision{The clustering obtained by LDA performs poorly, which is not surprising since LDA is not a clustering model for count data.} Finally, NMF and HTScluster also exhibit a strong stability to noise while clearly under-performing compared to other methods for this scenario.

\subsection{Model selection}
While the results above are encouraging for MMPCA, they are conducted with the true values $(Q^\star, K^\star) = (6, 4)$. This section evaluates the capacity of the ICL criterion proposed in Section \ref{sec:ModelSelection} for every value of $\lambda$ with $\epsilon=0$, since this corresponds to the true model. The results are shown in Table~\ref{table:ICL-AIC}, computed on $50$ datasets for each value of $\lambda $. It demonstrates a good performance for $\lambda=1$ and $0.85$, while seeming sensible to unbalanced clusters, as shown by the poor performance when $\lambda=0.7$. Interestingly, for $\lambda=0.7$, the criterion still selects $Q=4$ or $Q=5$, indicating that it could not capture smaller clusters, the high-dimensional setting with few data points for the smallest cluster complicating the asymptotic in approximations.
\begin{table}[ht!]
	\centering
	\begin{minipage}{.47\textwidth} \centering
		\begin{tabular}{|c|rrgr|}
			\hline
			Q \textbackslash K & 2 & 3 & 4 & 5 \\
			\hline
			2 & 0 & 0 & 0 & 0 \\
			3 & 0 & 0 & 0 & 0 \\
			4 & 0 & 0 & 0 & 0 \\
			5 & 0 & 0 & 0 & 0 \\
			\rowcolor{Gray}
			6 & 0 & 0 & 100 & 0 \\
			7 & 0 & 0 & 0 & 0 \\
			8 & 0 & 0 & 0 & 0 \\
			\hline
		\end{tabular}
		\captionof*{table}{$\lambda = 1$}
	\end{minipage} \hfill
	\begin{minipage}{.47\textwidth} \centering
		\begin{tabular}{|c||rrgr|}
			\hline
			Q \textbackslash K & 2 & 3 & 4 & 5 \\
			\hline
			2 & 0 & 0 & 0 & 0 \\ 
			3 & 0 & 0 & 0 & 0 \\ 
			4 & 0 & 0 & 0 & 0 \\ 
			5 & 0 & 0 & 2 & 0 \\ 	
			\rowcolor{Gray}
			6 & 0 & 0 & 98 & 0 \\ 
			7 & 0 & 0 & 0 & 0 \\ 
			8 & 0 & 0 & 0 & 0 \\ 
			\hline
		\end{tabular}
		\captionof*{table}{$\lambda = 0.85$}
	\end{minipage} \hfill
	\begin{minipage}{.47\textwidth} \centering
		\begin{tabular}{|c||rrgr|}
			\hline
			Q \textbackslash K & 2 & 3 & 4 & 5 \\
			\hline
			2 & 0 & 0 & 0 & 0 \\ 
			3 & 0 & 28 & 0 & 0 \\ 
			4 & 0 & 50 & 8 & 0 \\ 
			5 & 0 & 0 & 6 & 0 \\ 
			\rowcolor{Gray}
			6 & 0 & 0 & 8 & 0 \\ 
			7 & 0 & 0 & 0 & 0 \\ 
			8 & 0 & 0 & 0 & 0 \\ 
			\hline
		\end{tabular}
		\captionof*{table}{$\lambda = 0.7$}
	\end{minipage} \hfill
	\caption{Percentage of correct selections with ICL on $50$ simulated datasets. The actual number of cluster and topics are $Q^\star=6$ and $K^\star=4$.}
	\label{table:ICL-AIC}
\end{table}

\subsection{Sensitivity to sample size}
\begin{figure}[!ht]
	\centering
	\includegraphics[width=\textwidth]{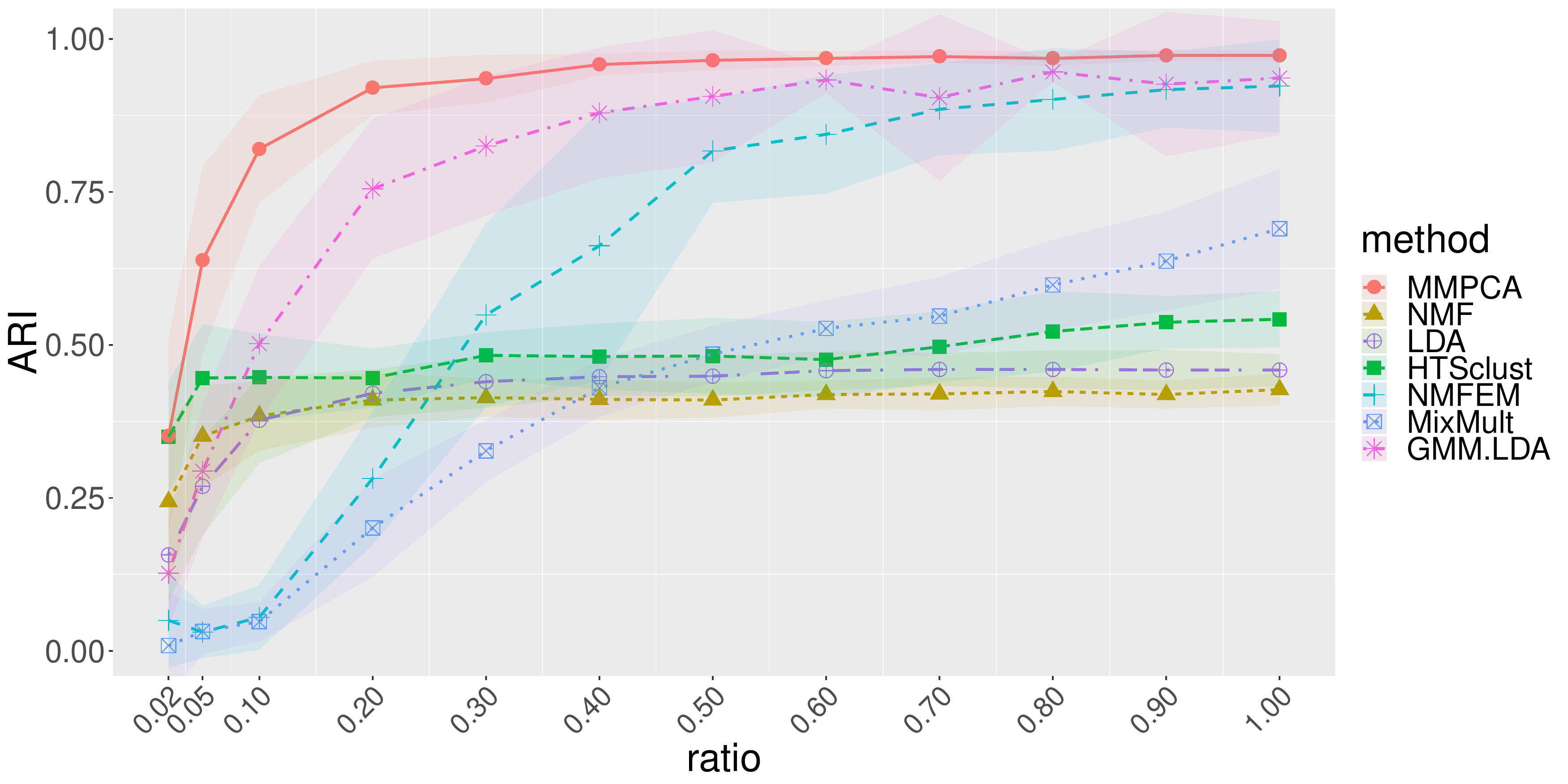}
	\caption{Stability with respect to sample size.}
	\label{fig:NonVRatio}
\end{figure}

This last experiment aims at comparing the sensibility of every methods to the dimensionality of the problem. Keeping the setting of Section \ref{subsec:RobustnessToNoise}, with $\lambda=0.85$ and $\epsilon=0.2$, \revision{$50$ datasets are simulated} with an increasing sample size. Results are shown in Figure \ref{fig:NonVRatio}, in term of the $N/V$ ratio. MMPCA clearly demonstrates a great stability beyond $N/V=0.1$, while \verb|GMM.LDA| seems to be more sensitive, even at large sample sizes as the error bars demonstrate. It also indicates that NMFEM can perform well in this experimental setting, which was expected, although it still needs far more observations than the aforementioned methods to reach the same performance. Basic mixture of multinomials also present some amelioration with an increased sample size, yet still suffering from the high dimensionality of the problem. As for NMF and HTSclust, they present a remarkable stability in this scenario, not seeming to benefit from the increasing number of observations.

\subsection{Computational complexity}

\begin{figure}[!h]
	\centering
	\includegraphics[width=\textwidth]{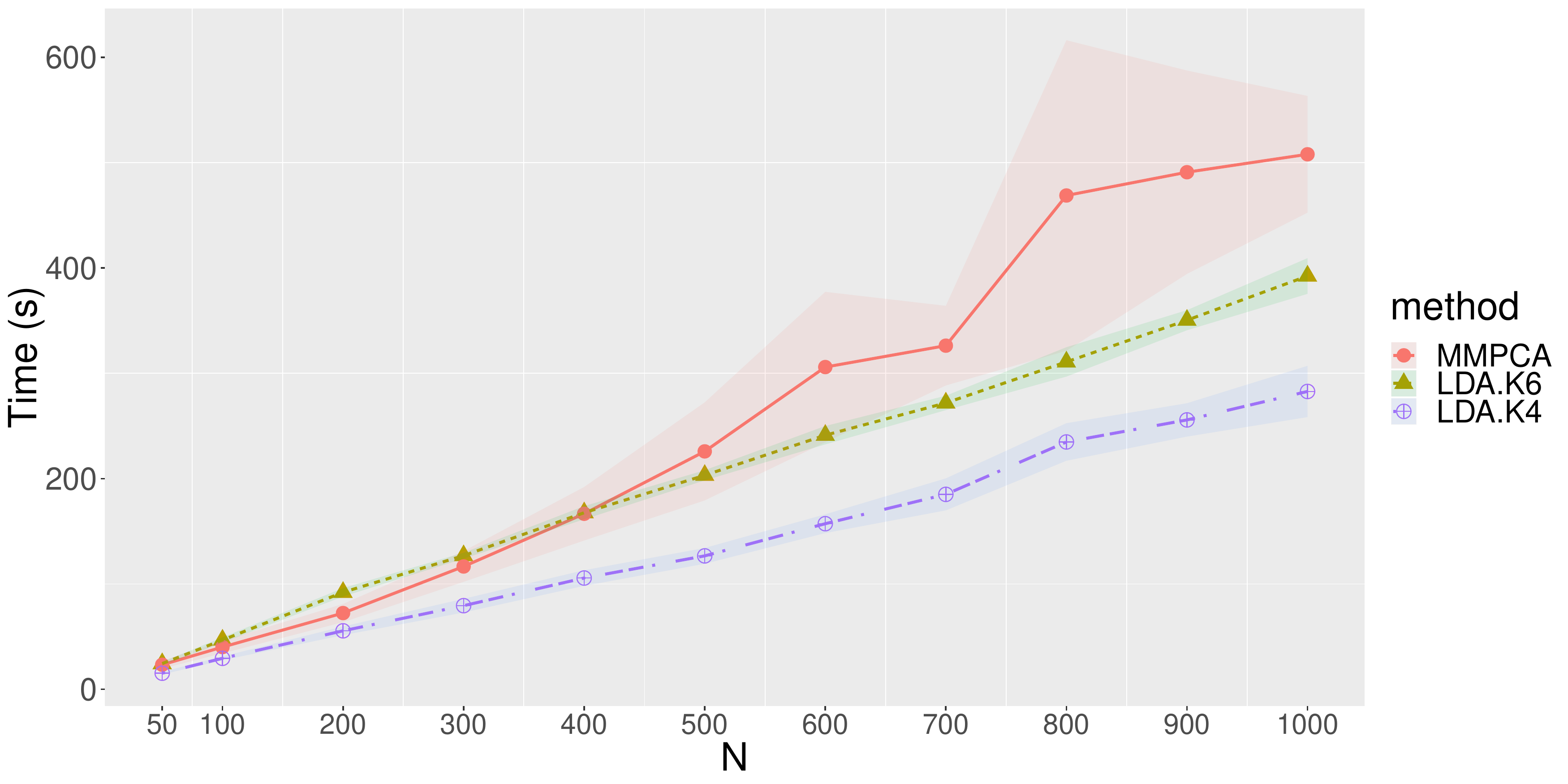}
	\caption{\revisiontwo{Mean computational time for $10$ runs of Algorithm~\ref{alg:BranchAndBound}, LDA with $K=6$ and $K=4$ topics. The number of observations is increasing while the dimension is fixed to $V=915$. The simulation setting is the same as the one in Section \ref{subsec:RobustnessToNoise} with $\epsilon=0$ and $\lambda=1$.}}
	\label{fig:ComputationalTime}
\end{figure}

\revisiontwo{Finally, Figure~\ref{fig:ComputationalTime} shows the computational time of Algorithm~\ref{alg:BranchAndBound} for increasing values of $N \in \{50, 100, \ldots, 1000 \}$, and for $Q=6$, $K=4$ and $V=915$. As we can see, Algorithm~\ref{alg:BranchAndBound} exhibits a linear growth with $N$, as discussed in Section~\ref{subsec:Complexity}. Moreover, the figure shows the complexity of running LDA with $K=6$ and $K=4$ topics. As we can see, relying on LDA.K6 for clustering, on LDA.K4 for topic modeling, or both at the same time, induces computational times of the same order of magnitude as Algorithm~\ref{alg:BranchAndBound}.}

\section{Applications to the clustering of anatomopathological reports}
\label{sec:RealData}

With 58,000 new cases in 2018 in France \citep{defossez2019estimations}, breast cancer is the most common malignant disease in women. Earlier diagnosis and better adjuvant therapy have substantially improved patient outcome. The pathologist establishes the diagnosis and provides prognostic and predictive factors of response to treatments. This is done by observing microscope slides of biological samples from both core needle biopsies and surgical specimens. Indeed, the microscopical aspect of cellular constituents and architecture are a fundamental part of diagnosis. Thus, information such as the histological type of the lesion \citep{lakhani2012classification}, the histopathological grading \citep{ellis2006histologic}, or molecular classification \citep{sorlie2003repeated}, are recorded in medical reports. The latter are heterogeneous, unstructured textual data, varying both with the pathologist writing style and with the change in medical conventions throughout the time. Although we have access to the pathologist conclusion on the lesion type, \textit{i.e.} the label, it is of interest to perform a deeper analysis to understand the variety and richness of information present.

\begin{figure}[!ht]
	\centering
	\includegraphics[scale=0.45]{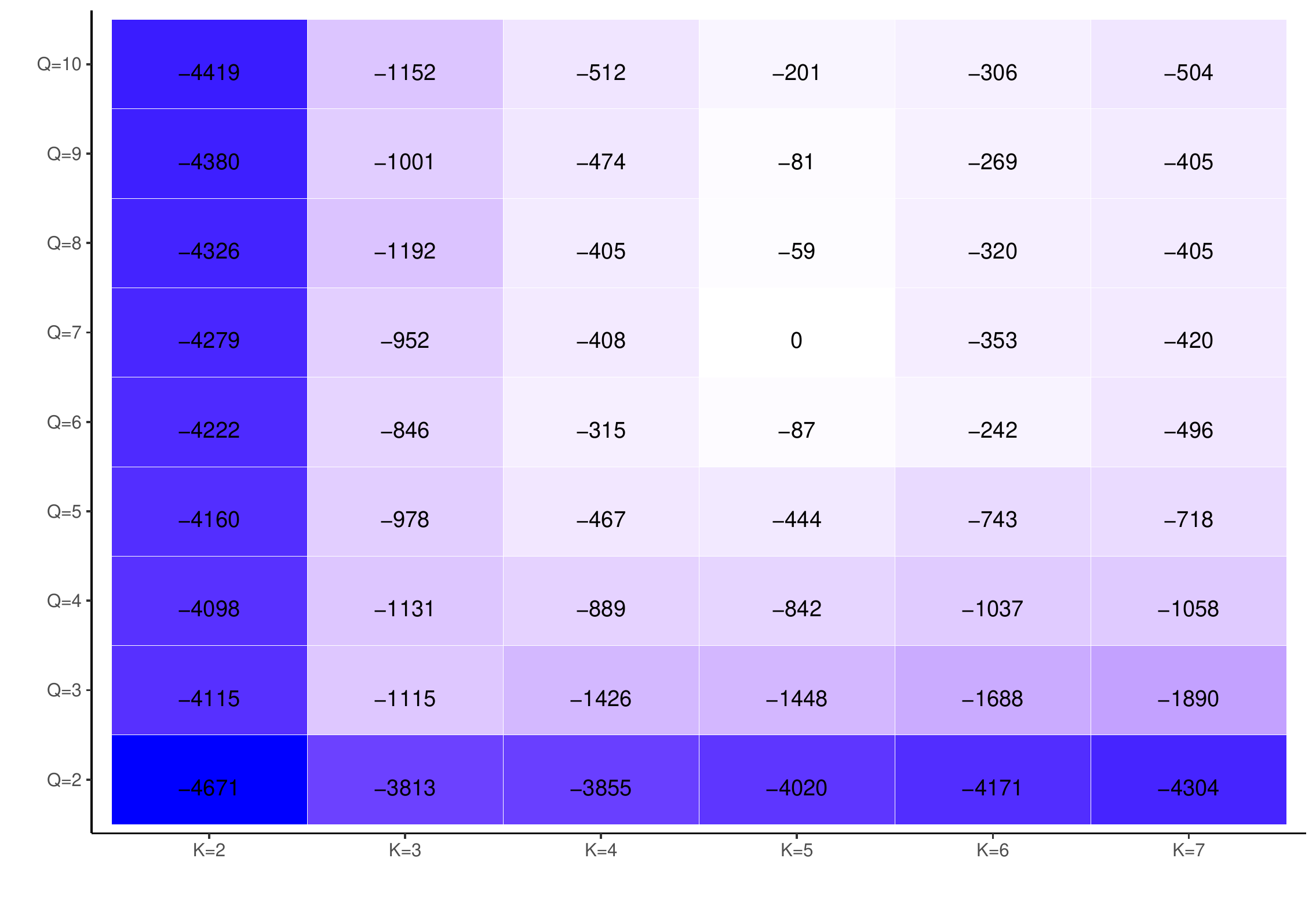}
	\caption{Model selection for MMPCA on the Curie anatomopathological reports datasets. The ICL criterion values are displayed with the maximum value subtracted for visualization purpose.}
	\label{fig:ICLcurie}
\end{figure}

The dataset considered here consists in about $900$ medical reports from the anatomopathological service of Institut Curie, a French hospital specialized in Cancer treatment. These reports describes histological lesions in tissues sampled from core needle breast biopsy. The lesions considered can be of two types: either benign, meaning there is no need for a medical care, or malignant lesion requiring specific care such as surgery, chemotherapy, and/or radiotherapy. World health organization classification of tumors of the breast divides malignant breast carcinomas in several types, including two main sub-categories \citep{lakhani2012classification}: non special type (NST, ex ductal) and lobular. In this study, only these two sub-types of invasive cancer are considered. Removing the conclusion from all documents, we only keep the descriptive part, and are interested in clustering those  anatomopathological reports to understand the information present in them.  For this, Algorithm \ref{alg:BranchAndBound} was run with $Q=2,\ldots,10$ and $K=2,\ldots, 7$ on a document-term matrix consisting of unigrams and a short hand designed word list. The vocabulary size is $302$ and the ICL criterion of Proposition \ref{prop:ICL} displayed in Fig. \ref{fig:ICLcurie} chose $Q=7$ clusters and $K=5$ topics.

In order to make a qualitative analysis of the results, Table \ref{table:resTextOnly} shows the labels repartition along clusters. The algorithm have found a clear separation between benign lesions in cluster $4$, lobular invasive carcinoma in cluster $1$, and NST invasive carcinoma splitted in the 5 smaller clusters. Observing the three NST documents in Cluster $4$ revealed that they focus a lot on describing benign lesions with minor invasive ones, thus explaining their clustering. Moreover, the smaller NST clusters are quite interesting since we recover some of the known prognostic and predictive factors of carcinomatous lesions. Indeed, cluster $5$ is the biggest cluster and correspond to high-grade invasive NST carcinoma which is expected. Cluster $7$ contains a lot of description of the stroma, which is known to have a major impact on response to the chemotherapy and patient outcome. As for the architecture aspect, Cluster $3$ and $6$ contains reports with well-differentiated architectures for the former and undifferentiated for the latter, implying a higher level of malignity. When looking at Cluster $2$, we may see that there is a lot of microcalcifications and in-situ\footnote{In-situ cancers are pre-invasive lesions that get their name from the fact that they have not yet started to spread. Invasive cancer tissues can contain both invasive and in-situ lesions in the same slide.} cancerous lesions in the reports descriptions. This can be explained by the fact that almost all samples present in this cluster came from a particular type of breast biopsy: macrobiopsy. These are almost exclusively used to search for cancerous lesion after the detection of microcalcifications in a breast mammography. Indeed, microcalcifications are considered as suspect in the development of cancerous tumors, especially the in-situ NST ones. This is interesting to know that we can recover information such as the type of medical exam from the description of tumorous lesions, when it does not appear in the text . 
\begin{table}[ht]
	
	\centering
	
	\begin{tabular}{rccc}
		
		\hline
		
		& Benign & Non special type carcinoma (ex ductal) & Lobular carcinoma \\
		
		\hline
		
		1 &   0 &   0 &  43 \\
		
		2 &   1 &  31 &   1 \\
		
		3 &   0 & 106 &   0 \\
		
		4 & 231 &   3 &   0 \\
		
		5 &   0 & 211 &   0 \\
		
		6 &   0 & 126 &   0 \\
		
		7 &   0 & 113 &   0 \\
		
		\hline
		
	\end{tabular}
	\caption{Confusion matrix of document label along cluster.}
	\label{table:resTextOnly}
	
\end{table}

\begin{figure}[!ht]
	\centering
	\includegraphics[scale=0.45]{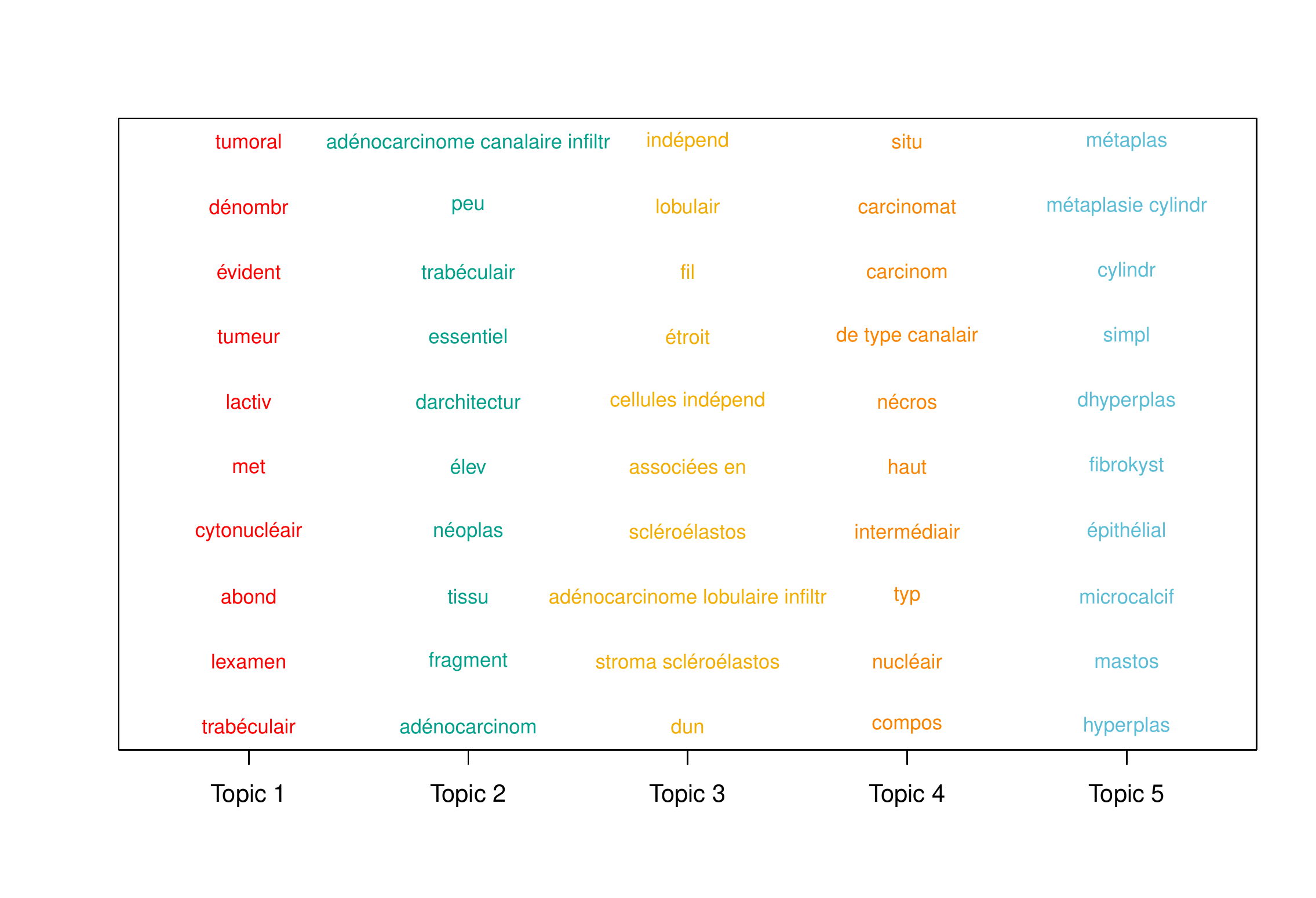}
	\caption{Most probable words per topics estimated on the aggregated Document Term Matrix.}
	\label{fig:Beta}
\end{figure}

Making use of the property described in Proposition \ref{prop:MetaDocConstruction}, we estimate the topic matrix $\beta$ and the cluster topic proportions $\theta$ on the $7$ meta-documents aggregated according to the final clustering. The variational estimates of all $\theta_q$, consisting of the normalized $\gamma_q$, is given in Table \ref{table:thetaQ}, while the most probable words per topic are shown in Figure \ref{fig:Beta}. The topic analysis provide a deeper insight and concordant results with the qualitative analysis above. 

\begin{itemize}[leftmargin=*, itemsep=10pt]
	\item[]\textbf{Topic 1.} This topic focus on general descriptive aspects of a tumor. In particular, words like "tumoral", "tumor", or "cytonuclear" are commonly used in medical reports when describing a tumor lesion. A word like "abundant" is related to stroma description, which explains why Cluster $7$ is peaked toward this topic. 
	
	\item[]\textbf{Topic 2.}  With keywords like "invasive ductal carinoma" corresponding to the lesion type and "poorly", "high" corresponding to the histopathological grading of the tumor \citep{ellis2006histologic}, this topic correspond to high-grade invasive ductal carinoma. Interestingly, Cluster 5 is completely peaked towards topic 2, and the analysis of the grade reveals that most of them are from intermediate to high.  
	
	\item[]\textbf{Topic 3.}  The keywords "independant cells" and "fibro-elastic stroma" are commonly used to describe "invasive lobular carcinoma" lesion. As expected, Cluster $1$ is entirely peaked toward this topic since it contains all invasive lobular carcinoma. 
	
	\item[]\textbf{Topic 4.}  Containing some keywords like "in situ", "high", "intermediate" or "necrosis", this topic is clearly related to the lexical field of in-situ lesions that can be associated with invasive cancer. We can see that Cluster $2$, $3$ and $6$ are associated to this topic. It was known for Cluster $2$ since it involves microcalcifications, however it brings some more information about the two other clusters.
	
	\item[]\textbf{Topic 5.} This topic is characteristic of the benign lesions lexical field. The keywords "cylindric metaplasia", "fibrocystic" or "simple" are related to benign breast lesions that are all grouped inside Cluster $4$. It also contains "microcalcification" which is characteristic of Cluster $2$ as explained above.
	
\end{itemize}
\begin{table}
	\centering
	\begin{tabular}{|r|ccccc|}
		\hline
		& Topic1 & Topic2 &Topic3 &  Topic4 & Topic5 \\
		\hline
		$\theta_1$ &   0.00&   0.01&   \textbf{0.98}&   0.00&   0.00 \\
		\hline
		$\theta_2$ &   0.19&   0.11&   0.04&   0.38&   0.29 \\
		\hline
		$\theta_3 $&   0.13&   0.09&   0.01&   \textbf{0.76}&   0.00 \\
		\hline
		$\theta_4$ &   0.01&   0.00&   0.01&   0.01&   \textbf{0.97} \\
		\hline
		$\theta_5$ &   0.00&   \textbf{1.00}&   0.00&   0.00&   0.00 \\
		\hline
		$\theta_6$ &   0.05&   \textbf{0.65}&   0.03&   0.26&   0.01 \\
		\hline
		$\theta_7$ &  \textbf{ 0.74}&  0.12&   0.03&   0.11&   0.00\\
		\hline
	\end{tabular}
	\caption{The matrix of estimated $(\theta_q)_{1, \ldots, 7}$. The topics are associated to those described in Figure \ref{fig:Beta}.}
	\label{table:thetaQ}
\end{table}

\section{Conclusion}
\label{sec:Conclusion}
\revision{In this work, we introduced a new algorithm for the clustering of count data based on a mixture of MPCA distributions, allowing to associate the dimension reduction aspect of topic modeling with model based clustering. The methodology maximizes a variational bound of an integrated classification likelihood of the model in a greedy fashion, handling both parameter inference and discrete optimization with respect to the partition. In addition, an ICL-like model selection criterion was proposed to select the number of clusters and topics.} Experiments on simulated data were used to assess the interest of the proposed approach, its performances comparing favorably with other methods in different scenarios. Notably, a real data application in medical report clustering illustrated the capacity to unveil some relevant structure from count data.

\section*{Acknowledgements}
	This work was supported by a DIM Math Innov grant from R\'{e}gion Ile-de-France. \revision{This work has also been supported by the French government through the 3IA Côte d'Azur Investments in the Future project managed  by the National Research Agency (ANR) with the reference number ANR-19-P3IA-0002. We are thankful for the support from fédération F2PM, CNRS FR 2036, Paris.} Finally, we would like to thank the anonymous reviewers for their helpful comments which contributed to improve the paper.

\appendix
\section{Proofs}
\subsection{Constructing meta-observation }
\label{appendix:PreuveMetadoc}
\begin{proof}[Proof of Proposition \ref{prop:MetaDocConstruction}]
	\begin{align*}
		\p(\W, \theta\mid Y, \, \beta) & = \p(\theta) \times \p(\W \mid \theta, Y) , \nonumber \\
		& = \prod_{q^\prime} \p(\theta_{q^\prime}) \times \prod_i \prod_q  \prod_n \mathcal{M}_V(w_{in}, \,  1 , \,\beta \theta_q)^{Y_{iq}} \nonumber, \\
		& = \prod_q \p(\theta_q) \prod_i \prod_v \prod_n (\beta_{v,\cdot} \theta_q)^{ Y_{iq} w_{inv}} \nonumber ,\\
		& = \prod_q \p(\theta_q)  \prod_v \prod_i (\beta_{v,\cdot} \theta_q)^{ Y_{iq} x_{iv}} \nonumber ,\\
		& = \prod_q \p(\theta_q) \prod_v (\beta_{v,\cdot} \theta_q)^{\sum_i Y_{iq} x_{iv}} ,
	\end{align*}
	since $x_{iv} = \sum_n w_{inv}$. Then, put
	\[\metadoc_q(Y) = \sum_{i=1}^N Y_{iq} x_{i}\,\]
	and this completes the proof of Proposition~\ref{prop:MetaDocConstruction}. 
\end{proof}

\subsection{Derivation of the lower bound}
\label{appendix:PreuveDecomposeBound}
\begin{proof}[Lower bound and Proposition \ref{prop:DecomposeBound}]
	The bound of Equation \eqref{eq:CELBO} follows from standard derivation of the evidence lower bound in variational inference. Since the $\log$ is concave, by Jensen inequality:
	\begin{align*}
		\log \p(\W, Y \mid \pi, \beta) &= \log \sum_Z \int_{\theta}  \p(\W, Y, \theta, Z \mid \pi, \beta) \dif \theta ,\\
		&= \log \sum_Z \int_{\theta} \frac{\p(\W, Y, \theta, Z \mid \pi, \beta)}{\q(Z, \theta ) } \q(Z, \theta )  \dif \theta ,\\
		&= \log \left( \mathbb{E}_{\q}\left[  \frac{\p(\W, Y, Z, \theta \mid \pi, \beta)}{\q(Z,\theta )}\right] \right)\\
		&\geq \mathbb{E}_{\q}\left[ \log \frac{\p(\W, Y, Z, \theta \mid \pi, \beta)}{\q(Z,\theta)}\right],\\
		&:= \mathcal{L}(\q(\cdot); \, \pi, \beta, \Y) .
	\end{align*}
	Moreover, the difference between the classification log-likelihood and its bound is exactly the KL divergence between approximate posterior $\q(\cdot)$ and the true one:
	\begin{align*}
		\log \p(\W, Y \mid \pi, \beta) - \mathcal{L}(\q(\cdot); \, \pi, \beta, \Y) &=  - \mathbb{E}_{\q}\left[ \log \frac{\p(Z, \theta \mid \W, \Y, \pi, \beta)}{\q(Z,\theta )}\right] .
	\end{align*}
	Furthermore, the complete expression is given in Proposition \ref{prop:DecomposeBound} as:
	\begin{align*}
		\mathcal{L}(\q(\cdot); \, \pi, \beta, \Y) =& \underbrace{\mathbb{E}_{\q}\left[ \log \p (\W, \Z, \theta \mid \Y, \beta)\right] - \mathbb{E}_{\q}\left[\log \q(Z, \theta) \right]}\limits_{\mathcal{J}_{\textrm{LDA}}} + \log \p (\Y \mid \pi) , \nonumber \\
		=& \sum_{q=1}^Q\mathcal{J}_{\textrm{LDA}}^{(q)}( \q;\,  \beta, \metadoc_q(Y)) + \sum_{q=1}^Q \sum_{i=1}^{N} Y_{iq}\log(\pi_q)  , \nonumber \\
	\end{align*}
	where
	\begin{align}
		\label{eq:detailedCELBO}
		\mathcal{J}_{\textrm{LDA}}^{(q)}( \q;\,  \beta, \metadoc_q(Y)) = 
		&  \log \Gamma(\tsum_{k=1}^{K} \alpha_k) - \sum_{k=1}^{K}\log \Gamma(\alpha_k)  \nonumber \\
		& + \sum_{k=1}^{K} (\alpha_k - 1) (\psi(\gamma_{qk}) - \psi(\textstyle \sum_{l=1}^K \gamma_{ql})) \nonumber \\
		& +   \sum_{i=1}^N Y_{iq} \sum_{k=1}^K  \sum_{n=1}^{L_i} \phi_{ink} \left[ \psi(\gamma_{qk}) - \psi(\textstyle \sum_{l=1}^K \gamma_{ql}) + \sum_{v=1}^{V} w_{inv} \log(\beta_{vk})\right]  \nonumber \\
		& - \log \Gamma(\tsum_{k=1}^{K} \gamma_{qk}) - \sum_{k=1}^{K}\log \Gamma(\gamma_{qk})   \\
		& - \sum_{k=1}^{K} (\gamma_{qk} - 1) (\psi(\gamma_{qk}) - \psi(\textstyle \sum_{l=1}^K \gamma_{ql})) \nonumber \\
		& -   \sum_{k=1}^K  (\gamma_{qk} -1) (\psi(\gamma_{qk}) - \psi(\textstyle \sum_{l=1}^K \gamma_{ql}))  \nonumber \\
		& - \sum_{i=1}^{N} Y_{iq} \sum_{n=1}^{L_i} \phi_{ink} \log(\phi_{ink}) \nonumber .
	\end{align} 
	
\end{proof}

\subsection{Optimization of $\q(\Z)$}
\label{appendix:PreuvePhi}
\begin{proof}[Proof of Proposition \ref{prop:PhiUpdate}]
	A classical result about mean field inference, see \citet{blei2017variational}, states that at the optimum, considering all other distributions fixed:
	\begin{align}	
		\log \q(z_ {in}) &= \mathbb{E}_{\Z^{ \setminus i, n}, \theta} \left[ \log \p (\W, \Z, \theta \mid \Y )\right] + \const, \nonumber
	\end{align}
	where the expectation is taken with respect to all $\Z$ except $z_{in}$, and to all $\theta$, assuming $(\Z, \theta) \sim \q$. Developing the latter leads to: 
	\begin{align}
		\label{eq:OptimumPhi}	
		\log \q(z_ {in}) &=  \sum_{k=1}^{K} z_{ink} \left[\sum_{v=1}^{V} w_{inv} \log(\beta_{vk}) + \sum_{q=1}^Q Y_{iq} \left\{\psi(\gamma_{qk}) - \psi(\textstyle \sum_{l=1}^K \gamma_{ql}) \right\}   \right]+ \const.
	\end{align}
	Equation \eqref{eq:OptimumPhi} characterizes the log density of a multinomial:
	\[
	\q(z_{in}) = \mathcal{M}_K(z_{in}; \, 1, \,\phi_{in} = (\phi_{in1}, \ldots, \phi_{inK})),
	\]
	where the quantity inside brackets represents the logarithm of the parameter, modulo the normalizing constant. Hence,
	\[
	\forall k, \quad \phi_{ink} \propto \left( \prod_{v=1}^V \beta_{vk}^{w_{inv}} \right) \, \prod_{q=1}^Q \exp\left\{ \psi(\gamma_{qk}) - \psi\left(\textstyle \sum_{l=1}^K \gamma_{ql}\right)\right\}^{Y_{iq}} .
	\]
\end{proof}
\subsection{Optimization of $\q(\theta)$}
\label{appendix:PreuveGamma}
\begin{proof}[Proof of Proposition \ref{prop:GammaUpdate}]
	With the same reasoning, the optimal form of $\q(\theta)$ is:
	\begin{align}
		\label{eq:PreuveGamma}
		\log \q(\theta)
		&= \mathbb{E}_{\Z}\left[ \p (\W, \Z, \theta \mid \Y ) \right] \, + \, \const\nonumber , \\
		&= \sum_{q=1}^{Q} \left[\sum_{k=1}^{K} (\alpha_k - 1) \log(\theta_{qk}) + \sum_{i=1}^{N} Y_{iq} \sum_{n=1}^{L_i}   \sum_{k=1}^{K} \phi_{ink} \log(\theta_{qk}) \right] + \, \const , \nonumber \\
		&= \sum_{q=1}^{Q}\sum_{k=1}^{K} \left[\alpha_k + \sum_{i=1}^{N} Y_{iq} \sum_{n=1}^{L_i} \phi_{ink} - 1 \right] \log(\theta_{qk}) \, + \, \const .
	\end{align}
	Once again, a specific functional form appears as the log of a product of $Q$ independent Dirichlet densities. Then,
	\[
	\q(\theta) = \prod_{q=1}^{Q} \Dir_K\left(\theta_q; \, \gamma_q=(\gamma_{q1}, \ldots, \gamma_{qK})\right) ,
	\]
	with the Dirichlet parameters inside the brackets of Equation \eqref{eq:PreuveGamma}:
	\[
	\forall (q,k), \quad \gamma_{qk} = \alpha_k + \sum_{i=1}^{N} Y_{iq}\sum_{n=1}^{L_i} \phi_{ink} .
	\]
\end{proof}

\subsection{Optimization of $\beta$}
\label{appendix:PreuveBeta}
\begin{proof}[Proof of Proposition \ref{prop:BetaAndPi} (I)] This a constrained maximization problem with $K$ constraints $\sum_{v=1}^{V} \beta_{vk} = 1$.  Isolating terms of Equation \eqref{eq:detailedCELBO} depending on $\beta$, and denoting constraints multipliers as $(\lambda_k)_k$, the Lagrangian can be written:
	\begin{align*}
		f(\beta, \lambda) = & \sum_{q=1}^{Q} \sum_{i=1}^{N} Y_{iq} \sum_{n=1}^{L_i} \sum_{v=1}^{V} \phi_{ink} w_{inv} \log(\beta_{vk}) + \sum_{k=1}^{K} \lambda_k (\beta_{vk} - 1) , \nonumber \\
		= & \sum_{i=1}^{N} \sum_{n=1}^{L_i} \sum_{v=1}^{V} \phi_{ink} w_{inv} \log(\beta_{vk}) + \sum_{k=1}^{K} \lambda_k (\beta_{vk} - 1) .
	\end{align*}
	Setting its derivative to $0$ leaves:
	\[
	\beta_{vk} \propto  \sum_{i=1}^{N} \sum_{n=1}^{L_i} \phi_{ink} \, w_{inv} .
	\]
	
\end{proof}

\subsection{Optimization of $\pi$}
\label{appendix:PreuvePi}
\begin{proof}[Proof of Proposition \ref{prop:BetaAndPi} (II)]
	The bound depends on $\pi$ only through its clustering term:
	\[
	\log \p(Y \mid \pi) = \sum_{i=1}^{N}\sum_{q=1}^{Q} Y_{iq} \log(\pi_q)  .
	\]
	Once again, this is a constrained optimization problem, and, introducing the Lagrange multiplier $\lambda$ associated to the constraint $\tsum_{q=1}^{Q} \pi_q = 1$, we get:
	\[
	\sum_{q=1}^{Q} \sum_{i=1}^{N} Y_{iq} \log(\pi_q) + \lambda (\tsum_{q=1}^{Q} \pi_q - 1) .
	\]
	Setting the derivative with respect to $\pi_q$ to $0$, we get:
	\[
	\pi_q = \frac{\sum_{i=1}^{N} Y_{iq}}{N} .
	\]
\end{proof}

\subsection{Model selection}
\label{appendix:PreuveICL}

\begin{proof}[Proof of Proposition \ref{prop:ICL}]
	Assuming that the parameters $(\pi, \beta)$ follows a prior distribution that factorizes as follow:
	\begin{equation}
	\p(\pi, \beta \mid Q, K)  = \p(\pi \mid Q, \eta) \, \p(\beta \mid K),
	\end{equation}
	where
	\begin{equation}
	\p(\pi \mid Q, \eta) =\mathcal{D}_K(\pi ; \, \eta \mathbf{1}_Q) .
	\end{equation}
	Then, the classification log-likelihood is written:
	\begin{eqnarray}
	\label{eq:integratedcompletellhood}
	\log\p(\W, \Y \mid Q, K) & = &  \log \int_{\pi} \int_{\beta}\p(\W,Y, \beta, \pi \mid Q, K) \, \dif \pi\, \dif \beta \nonumber\\
	&=& \log \int_{\pi} \int_{\beta}\p(\W,Y \mid \beta, \pi , \, Q, K) \p(\pi \mid Q, \eta) \, \p(\beta \mid K) \, \dif \pi\, \dif \beta \nonumber \\
	& = & \log \int_{\pi} \p(Y \mid \pi)  \p(\pi \mid Q, \eta) \dif\pi \, \int_{\beta}\p(\W \mid Y, \beta, Q, K)  \p(\beta \mid K)   \dif\beta \nonumber \\
	&=& \log \int_{\pi} \p(Y \mid \pi)  \p(\pi \mid Q, \eta) \dif\pi  \nonumber\\
	& &   \qquad \qquad  \qquad  \qquad + \log \int_{\beta}\p(\W \mid Y, \beta, Q, K)  \p(\beta \mid K)  \dif\beta  .
	\end{eqnarray}
	The first term in Equation~\eqref{eq:integratedcompletellhood} is exact by Dirichlet-Multinomial conjugacy. Setting $\eta=\frac{1}{2}$ plus a Stirling approximation on the Gamma function as in \citet{daudin2008mixture} leads to:
	\begin{equation}
	\log \int_{\pi} \p(Y \mid \pi)  \p(\pi \mid Q, \eta) \dif\pi \approx \max\limits_{\pi} \log \p(Y \mid \pi, Q)  - \frac{Q-1}{2} \log(D) .
	\end{equation}
	As for the second term, a BIC-like approximation as in  \citet{bouveyron2018stochastic} gives:
	\[
	\log \int_{\beta}\p(\W \mid Y, \beta, Q, K)  \p(\beta \mid K)  \dif\beta \approx  \max\limits_{\beta} \log \p(\W \mid Y, \beta, Q, K) - \frac{K (V-1)}{2} \log(Q).
	\]
	In practice, $ \log \p(\W \mid Y, \beta, Q, K) $ is still intractable, hence we replace it by its variational approximation after convergence of the VEM, $\mathcal{J}^\star_{\textrm{LDA}}$, which is the sum of the meta-observations individual LDA-bounds detailed in Equation \eqref{eq:detailedCELBO} (different from $\mathcal{L}$). In the end, it gives the following criterion:
	\begin{eqnarray}
	\ICL(Q, K, Y,  \W) & = & \mathcal{J}^\star_{\textrm{LDA}}(\q; \, \beta, Y) - \frac{K (V-1)}{2} \log(Q)   \nonumber \\ 
	& &+ \max\limits_{\pi} \log \p(Y \mid \pi, Q)  - \frac{Q-1}{2} \log(D) .
	\end{eqnarray}
	Note that: 
	\[
	\max\limits_{\beta} \log \p(\W \mid Y, \beta, Q, K) + \max\limits_{\pi} \log \p(Y \mid \pi, Q) \approx \mathcal{L}^\star ,
	\]
	\emph{i.e.} the bound after Algorithm \ref{alg:BranchAndBound} converges.
	
\end{proof}

\bibliographystyle{apalike}
\bibliography{biblio_article}

\end{document}